\documentclass[%
reprint,
superscriptaddress,
showpacs,preprintnumbers,
bibnotes,
 amsmath,amssymb,
 aps,
 pre,
floatfix,
]{revtex4-1}

\UseRawInputEncoding
\usepackage{natbib}
\usepackage{graphicx}
\usepackage{dcolumn}
\usepackage{amsmath,amsthm}
\usepackage{amsfonts}
\usepackage{amssymb}
\usepackage{bbold}
\usepackage{times}
\usepackage{mathrsfs}
\usepackage{color}
\usepackage{gensymb}
\usepackage{times}
\usepackage{bm}
\usepackage{stmaryrd}
\usepackage{comment}
\DeclareGraphicsExtensions{.PNG,.jpg,.pdf,.gif,.eps}
\usepackage{amsmath,amssymb,amsthm,amsfonts,eucal}
\usepackage[english]{babel}
\usepackage{mathtools, nccmath}
\usepackage{epstopdf}
\usepackage[top=0.75in,left=0.75in,right=0.75in,bottom=0.75in]{geometry}
\usepackage{float}
\usepackage[caption=false]{subfig}
\newcommand{\subfigimg}[3][,]{%
  \setbox1=\hbox{\includegraphics[#1]{#3}}
  \leavevmode\rlap{\usebox1}
  \rlap{\hspace*{-0.05cm}\raisebox{\dimexpr\ht1+0.3\baselineskip}{#2}}
  \phantom{\usebox1}
}
\newcommand{\e}{\varepsilon}

\usepackage{longtable}

\graphicspath{{Figures/}}

\def\beq{\begin{equation}}
\def\eeq{\end{equation}}

\begin{document}


\title{Transition fronts and their universality classes}

\author{N. Gorbushin}
\affiliation{\it  PMMH, CNRS -- UMR 7636, CNRS, ESPCI Paris, PSL Research
University, 10 rue Vauquelin, 75005 Paris, France}

\author{A. Vainchtein}
\affiliation{\it Department of Mathematics, University of Pittsburgh, Pittsburgh, Pennsylvania 15260, USA}

\author{L. Truskinovsky}
\affiliation{\it  PMMH, CNRS -- UMR 7636, CNRS, ESPCI Paris, PSL Research
University, 10 rue Vauquelin, 75005 Paris, France}

\begin{abstract}
Steadily moving transition (switching) fronts, bringing local transformation, symmetry breaking  or collapse,  are among the most important dynamic coherent structures. The nonlinear mechanical waves of this type play a major role in many modern applications involving the transmission of mechanical information in systems  ranging from crystal lattices and metamaterials to macroscopic civil engineering structures. While many different classes of such dynamic fronts  are known, the interrelation between them remains obscure. Here we consider a minimal prototypical mechanical system, the Fermi-Pasta-Ulam (FPU) chain with piecewise linear nonlinearity, and show that there are exactly three distinct classes of switching fronts, which differ fundamentally in how (and whether) they produce and transport oscillations. The fact that all three types of fronts could be obtained as explicit Wiener-Hopf solutions of the same discrete FPU problem, allows one to identify the exact mathematical origin of the particular features of each class.  To make the underlying Hamiltonian dynamics analytically transparent, we  construct  a minimal quasicontinuum approximation of the FPU model that captures all three classes of the  fronts and interrelation between them. This approximation is of higher order than conventional ones (KdV, Boussinesq) and involves mixed space-time derivatives. The proposed framework unifies previous attempts to classify the mechanical transition fronts as radiative, dispersive, topological or compressive and categorizes them instead as different types of dynamic lattice defects.
\end{abstract}

\date{\today}

\maketitle

\section{Introduction}
Transition fronts in discrete systems continue to attract a lot of attention because they represent examples of far-from equilibrium collective phenomena that emerge from the underlying many-body interactions. Interpreted as highly nonlinear coherent dynamic structures, such fronts play an important role in the energy transmission from macro to microscales. They are observed in both integrable and non-integrable Hamiltonian systems \cite{kamvissis1993toda,holian1978molecular}, can be topological or non-topological \cite{flytzanis1985kink,peyrard1986discreteness,deng2020nonlinear},
spreading or compact \cite{lowman2013fermionic}, compressive or undercompressive (non-Lax) \cite{hayes1999undercompressive},
stable or unstable \cite{an2018dispersive}. Together with solitons and breathers, they play a crucial role as building blocks in complex nonlinear wave patterns that emerge generically in mechanical systems ranging from crystals \cite{yasuda2017emergence,vattre2019continuum,baqer2020modulation} to  nanomechanical structures \cite{sato2006colloquium,raney2016stable,nadkarni2016unidirectional,nadkarni2014dynamics}.

The concept of transition fronts is equally relevant for the description of pattern formation \citep{beck2009snakes} and transport properties in nonmechanical dynamical systems, including coupled waveguide arrays \citep{ricketts2018electrical,fleischer2005spatial},
quantum systems  \citep{binder2000observation,chevriaux2006theory}, Bose-Einstein condensates  \cite{kevrekidis2007emergent,mossman2018dissipative,morsch2006dynamics,peotta2014quantum},
electronic liquids \citep{bettelheim2006nonlinear}, ultracold quantum gases \citep{kamchatnov2004dissipationless,hoefer2006dispersive},
rarefield plasma \citep{tran1977shocklike}, intense electron beams \citep{mo2013experimental}, liquid helium  \citep{rolley2007hydraulic},
and exciton polaritons \citep{dominici2015real}. In this paper we focus on mechanical switching fronts due to their the importance of their dynamics for the design of modern metamaterials \cite{yasuda2020transition,raney2016stable,kochmann2017exploiting,zhang2019programmable}.
The term  mechanical metamaterials is used here to describe high-contrast (soft-hard) composite structures with complex architecture at mesoscale. Characteristically, macroscopic properties of such structures are controlled more by the structural stability of the sub-elements then by their material properties \citep{clausen2015topology,bertoldi2017flexible,xia2019electrochemically,
hussein2014dynamics,christensen2015vibrant,chen2014metamaterials,kochmann2017exploiting,pishvar2020foundations}.
The use of additive manufacturing techniques opened a way to exploit various elastic instabilities embedded in the metamaterial response and to creatively guide them using applied deformation \citep{rafsanjani2016bistable,raney2016stable,chen2018harnessing}.
Dynamic effects  targeted by  various  metamaterial architectures include  mitigation  of  impact loadings, non-destructive detection of  inhomogeneities,  suppression or amplification of internal  instabilities, transmission, guiding and encryption of  mechanical information including the  enabling of logic operations,  dynamic unfolding of deployable structures, energy harvesting and even activating soft robotics
\citep{foehr2018spiral,nasrollahi2017nondestructive,
stawiarski2017fatigue,tan2014blast,shan2015multistable,
dorin2019vibration,kidambi2017energy,
wang2016tunable,zhang2019programmable,
deng2019propagation,yasuda2019origami,gorbushin2021peristalsis}.

One of the most interesting nonlinear dynamic effects which qualifies metamaterials as mesoscopic analogs of ordered solid-state materials, like ferroelectrics, ferromagnets and ferroelastics, is their ability to support moving transition fronts (analogs of domain boundaries), which enable the system to perform dynamic switching between different equilibrium states \cite{yang2016phase,frazier2017atomimetic,jin2020guided,zareei2020harnessing,kang2013buckling,
paulose2015selective,rafsanjani2019propagation,yasuda2020transition}. There is already a rich body of theoretical and experimental literature
devoted to the study of such dynamic snapping/switching  waves in mesoscopic mechanical
systems \cite{slepyan1984fracture,nadkarni2014dynamics,jin2020guided,manevich1994solitons,katz2018solitary}.
The ability to propagate transition fronts in metamaterials opens new ways towards potential
applications in  shape morphing, reconfigurable devices, mechanical logic, and controlled energy absorption \cite{rafsanjani2018kirigami,preston2019soft,chen2018harnessing,novelino2020untethered,deng2020pulse,fang2017asymmetric}. Analysis of low-dimensional model systems can serve as a guide for the structural design and optimization of the actual 3D mechanical systems.

Despite the  ubiquity of transition fronts in metamaterials, the relation between different \emph{classes} of such mobile  nonlinear dynamic structures  remains obscure.  In this paper we consider a well known prototypical system, the Fermi-Pasta-Ulam (FPU) model \cite{fermi1955studies,gallavotti2007fermi,berman2005fermi,chendjou2018fermi}
and present a unified description of the three main types of steady transition
fronts in this one-dimensional lattice, which we identify as \emph{subkinks}, \emph{shocks} and \emph{superkinks}. Various realizations of these archetypes have been previously  encountered in applications and  treated as unrelated:  subkinks  as  subsonic  phase  boundaries \cite{truskinovskii1987dynamics,truskinovsky2008dynamics,slepyan2001feeding},
shocks as classical supersonic  shock waves \cite{trofimov2010shocks, truskinovsky1993kinks} and superkinks as supersonic activity waves \cite{gorbushin2020supersonic,gorbushin2021peristalsis}.  They were first treated as disconnected  solutions of the FPU model in \cite{slepyan1984fracture,slepyan2012models}.  Some conceptual links  between subkinks and shocks have been previously established in \cite{truskinovsky1993kinks,trofimov2010shocks}, while superkinks remain a disconnected class of transition fronts \cite{iooss2000travelling,HerrmannRademacher10,Herrmann11,gorbushin2020supersonic}.

A unified description of all these transition fronts can be obtained if we use the simplest choice of nonlinearity  and  assume that the FPU interactions are piecewise linear. In fact, such interactions were already considered  in the original paper \cite{fermi1955studies} and have since been employed for the description of various dynamic nonlinear phenomena, e.g. \cite{atkinson1965motion, kresse2003mobility, kresse2004lattice, truskinovsky2005kinetics,slepyan2005transition, slepyan2012models}.

More specifically, we consider the Hamiltonian dynamics of a mass-spring chain with mass displacements $u_n(t)$ satisfying the infinite system of equations
\begin{equation}
\rho h\frac{d^2 u_n(t)}{dt^2}=\sigma\left(\frac{u_{n+1}-u_{n}}{h}\right)-\sigma\left(\frac{u_{n}-u_{n-1}}{h}\right).
\label{eq:EoM_displ}
\end{equation}
Here $h$ is the equilibrium distance between the masses $m=\rho h$, where $\rho$ is the mass density.
In terms of the strain variables
\[
\varepsilon_n(t)=\dfrac{u_{n+1}(t)-u_n(t)}{h},
\]
the equations become
\begin{equation}
\rho h^2\frac{d^2 \e_n(t)}{dt^2}=\sigma(\e_{n+1})-2\sigma(\e_n)+\sigma(\e_{n-1}).
\label{eq:EoM_strain}
\end{equation}
The  assumed  piecewise linear macroscopic  stress-strain relation can be written as
\beq
\sigma(\varepsilon)=\begin{cases} E_1\varepsilon, & \varepsilon<\varepsilon_c\\
                                  E_2\varepsilon-\sigma_0, & \varepsilon>\varepsilon_c,
                     \end{cases}
\label{eq:stress-strain}
\eeq
where  $\varepsilon_c$  is the critical (switching) strain, and $E_1$, $E_2$ are the elastic moduli in the two linear regimes. We assume that $E_2>E_1$, so that the two characterstic speeds  $c_{1,2}=\sqrt{ E_{1,2}/\rho}$
satisfy $c_2>c_1$. The corresponding
piecewise quadratic elastic energy density $\phi(\e)=\int \sigma(\e)d\e$ is continuous:
\[
\phi(\e)=\begin{cases} \frac{E_1}{2}\e^2,  & \e<\varepsilon_c, \\
                        \frac{E_2}{2}(\e^2-\varepsilon_c^2)-\sigma_0(\e-\varepsilon_c)+\frac{E_1}{2}\varepsilon_c^2, &  \e>\varepsilon_c.
                        \end{cases}
\]

Note that as the stress jump at the critical strain
\[
\Delta\sigma=\sigma(\e_c-0)-\sigma(\e_c+0)=\sigma_0-(E_2-E_1)\varepsilon_c
\]
varies from positive to negative values, we obtain two fundamentally different types of stress-strain curves. Thus, the elastic energy density $\phi(\e)$ is nonconvex when $\Delta\sigma>0$ and convex for $\Delta\sigma<0$. In the first case, the different branches of the stress-strain curve can be considered as different `phases' of the material, with spinodal region (where $\phi(\e)$ is concave in a smoother setting) represented by the single point $\varepsilon=\varepsilon_c$. In the second case ($\Delta\sigma<0$), the stress jump at $\varepsilon=\varepsilon_c$ is just a representation of the hardening-type nonlinearity, which is again concentrated at a single point. The advantage of the piecewise linear choice  for the stress-strain relation  is  the  possibility  to construct   the corresponding traveling wave solutions of the FPU problem explicitly  using the Wiener-Hopf (WH) transform technique \cite{slepyan2012models}.
While smoothening the constitutive response around the singular point $\varepsilon_c$ could make the model more realistic, sometimes  even without sacrificing much of analytical transparency \cite{vainchtein2010role,shiroky2017propagation,herrmann2013subsonic}, stronger nonlinearity is needed to capture such important physical effects as thermalization of the radiated phonons \cite{efendiev2010thermalization,benedito2020unfolding,blake2020dynamic}.
However, such generalization of the model, which will make its analytical treatment almost impossible without contributing  much to the classification of the transition fronts, is outside the scope of this paper.

To make the structure of the underlying  Hamiltonian  dynamics  clearly visible, we pose the problem of constructing the minimal quasicontinuum (QC) approximation of the FPU model capturing all three classes of the fronts.  The term quasicontinuum is used here in the sense that it is a continuum approximation of the discrete system, which is, however, not scale-free and carries a memory about the lattice discreteness \cite{truskinovsky2006quasicontinuum}. Our analysis shows  that the desired approximation must be necessarily of higher order than the conventional ones (KdV, classical `good' or `bad' Boussinesq) and should involve mixed space-time derivatives.  The obtained minimal QC model with desired properties can be viewed as a higher order version of the `good'  Boussinesq approximation \cite{christov2007boussinesq}. In contrast to the more conventional  approach of adding spatially nonlocal terms to the elastic energy \cite{kunin2012elastic,charlotte2008towards}, it introduces  the higher order derivatives into  the inertial part of the model (into the kinetic energy), as advocated earlier in \cite{charlotte2012lattice}.

The  proposed QC framework not only provides a transparent interpretation of the   three types of transition fronts as heteroclinic trajectories of different kinds in the phase space, but also helps to explain in physical terms    why some kinks are radiative (dissipative), while others are not, why some shocks are dispersive, while others are not, and why kinks are topological, while shocks are not. The comparison with the exact solutions of the discrete problem shows that, on both qualitative and quantitative levels, the relation between different classes of transition fronts is captured adequately by this minimal QC approximation.

It is important to mention that while the non-stationary (spreading) dispersive shock waves (DSW)  \cite{kamchatnov2019dispersive,chong2018dispersive,benzoni2021modulated} are not the focus of our study, which aims to classify steadily moving transition fronts, we show numerically that the DSWs replace the steady transition fronts in a subdomain of the parameter space. The adequacy of the QC approximation is corroborated by the fact that  the DSW  stability subdomains in discrete and QC models nearly overlap.

On a theoretical side, our approach unifies for the first time the previous attempts to classify the mechanical transition  fronts as radiative, dispersive, topological or compressive   and categorizes them instead in a unified framework  as fundamentally distinct types of dynamic lattice defects. The obtained analytical solutions  can be also used  in applications as a  guidance in  the design of new metamaterials exploiting  structural nonlinearity at the scale of the periodicity cell. For instance, our analysis  points to a particular   type of nonlinearity which should be used if the goal is the suppression of shock loading by channeling  the largest amount of energy  from macro to micro scales. It is also makes clear that a different type of nonlinearity must be engineered if the task is to transmit mechanical information with minimal losses. There is of course still a long way from our prototypical 1D designs to the construction of real 3D mesoscopic composite structures.

The rest of the paper is organized as follows. In Sec.~\ref{Continuum model} we formulate the classical continuum approximation of the discrete problem and  identify irreducible classes of transition fronts. Then in Sec.~\ref{Quasicontinuum model} we introduce a non-classical quasicontinuum  approximation of the  same discrete problem and construct explicit  solutions of the corresponding dispersive traveling wave problem describing all three distinct types of transition fronts.  In particular, we discuss the issues of solution admissibility in the piecewise linear model and the effective energy dissipation in this Hamiltonian framework  and present the results of direct numerical simulations that suggest stability of the obtained traveling waves. In Sec.~\ref{Discrete model} we construct an explicit traveling wave solution of the original discrete problem providing a unified description of all three types of fronts.  We  also present numerical simulations illustrating stability of the different types of transition fronts in various domains of the parameter space. In Sec.~\ref{Applications to metamaterials} we briefly mention potential applications of our results for the design of metamaterials. Summary of the results and concluding remarks can be found in Sec.~\ref{Conclusions}. Some asymptotic results are presented in the Appendix~\ref{sec:app}.

\section{Continuum model}
\label{Continuum model}
In our search of a unified description for the  different  types of transition fronts,  it is natural to start with  the classical continuum approximation of the original discrete model \eqref{eq:EoM_displ}. It can be obtained by taking a formal limit $h \to 0$ and replacing finite differences by the lowest order derivatives. Following \cite{blanc2002molecular}, we obtain the standard nonlinear wave equation, which can be usually represented as the
first-order system
\beq
\dfrac{\partial \varepsilon}{\partial t}= \dfrac{\partial v}{\partial x}, \quad
\rho \dfrac{\partial v}{\partial t} = \dfrac{\partial}{\partial x}\sigma(\varepsilon).
\label{eq:cont}
\eeq
Here $\varepsilon(x,t)=u_x$ and $v(x,t)=u_t$ are the strain and particle velocity, respectively. The system \eqref{eq:cont} has discontinuous solutions, which must satisfy the classical Rankine-Hugoniot (RH) conditions
\beq
\llbracket v \rrbracket + V\llbracket \varepsilon \rrbracket=0, \quad \rho
V\llbracket v \rrbracket+\llbracket \sigma(\varepsilon)\rrbracket=0,
\label{eq:RHconds}
\eeq
where  $V$ is the velocity of the jump discontinuity.  The notation  $\llbracket f \rrbracket\equiv f_+ - f_-$ will be used throughout the paper to describe the jump between the limiting values to the right and to the left of a discontinuity.

\begin{figure*}
  \begin{tabular}{@{}p{0.25\linewidth}@{\quad}p{0.25\linewidth}@{\quad}p{0.25\linewidth}@{}}
    \subfigimg[width=\linewidth]{(a)}{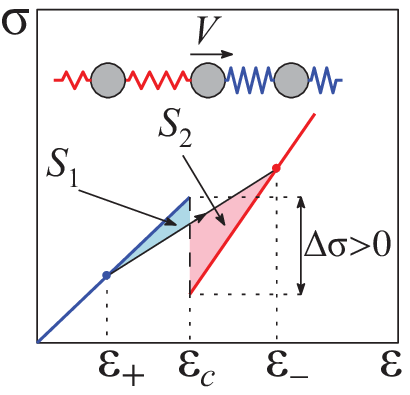} &
    \subfigimg[width=\linewidth]{(b)}{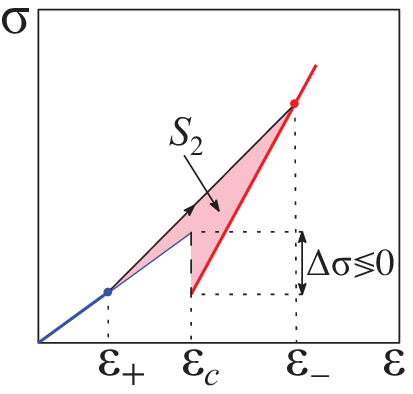} &
    \subfigimg[width=\linewidth]{(c)}{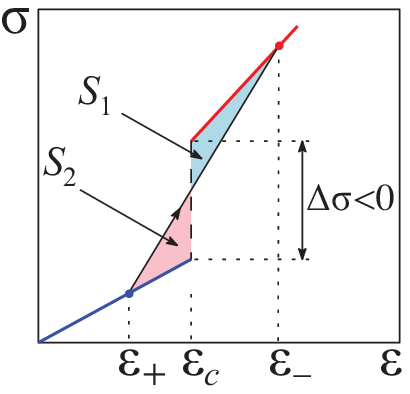}
  \end{tabular}
  \caption{Rayleigh lines connecting $(\e_{+},\sigma(\e_{+}))$ and $(\e_{-},\sigma(\e_{-}))$ with the slope $\rho V^2$ satisfying Eq.~\eqref{eq:RH} for three distinct types of traveling wave solutions: (a) subsonic kinks, $V<c_1<c_2$, (b) shocks, $c_1<V<c_2$, (c) supersonic kinks, $c_1<c_2<V$. The driving force is $G=S_2-S_1$, where $S_1$ (blue) and $S_2$ (pink) are the areas cut by the Rayleigh line from the stress-strain curve.}
\label{fig:stress_strain}
\end{figure*}

By changing the parameter $\Delta\sigma$ and varying independently the velocity of the jump discontinuity, we can obtain three fundamentally different types of steadily moving transition fronts shown schematically in Fig.~\ref{fig:stress_strain}.
Each  transition front connects a state $\varepsilon=\varepsilon_{+}$ in front with a state $\varepsilon=\varepsilon_{-}$ behind. Both of these states $\varepsilon_{\pm}$ belong to the stress-strain curve which is  piecewise linear, and to be nontrivial the transition front must connect the states on two sides of the singular point  $\varepsilon=\varepsilon_{c}$. The RH conditions state that the slope of the Rayleigh line connecting $(\e_{+},\sigma(\e_{+}))$ and $(\e_{-},\sigma(\e_{-}))$ is proportional to the square of the velocity $V$ of the front:
\begin{equation}
\sigma(\varepsilon_{+})-\sigma(\varepsilon_{-})=\rho V^2(\varepsilon_{+}-\varepsilon_{-}).
\label{eq:RH}
\end{equation}
The three different types of transition fronts are defined by the relation between their velocity $V$ and the characteristic velocities $c_1$ and $c_2$, which can be determined by comparing the slopes of the Rayleigh line and the corresponding linear regimes of the stress-strain curve. In what follows, we will refer to them as subkinks (subsonic kinks, $V<c_1<c_2$, panel (a) of Fig.~\ref{fig:stress_strain}), shocks (intersonic fronts, $c_1<V<c_2$, panel (b)) and  superkinks (supersonic kinks, $c_1<c_2<V$, panel (c)).

\subsection{Well posedness}
Note that there are five variables to be determined for each discontinuity: $v_{\pm}$, $\varepsilon_{\pm}$
and $V$. Two relations between these five unknowns are furnished by the RH conditions \eqref{eq:RHconds}. Fig.~\ref{fig:stress_strain} shows qualitatively  the fundamentally different relations of this type. Additional information can be obtained by solving the problem \eqref{eq:cont} using the method of characteristics. Due to the piecewise linear nature of the problem, two families of  characteristics  with velocities $\pm c_{1,2}$  can be defined on both  sides of the moving front.

\begin{figure}[h!]
  \centering
  \begin{tabular}{@{}p{0.3\linewidth}@{\quad}p{0.3\linewidth}@{\quad}p{0.3\linewidth}@{}}
    \subfigimg[scale=1.4 ]{(a)}{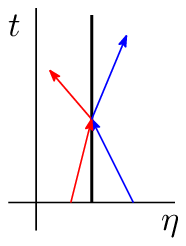} &
    \subfigimg[scale=1.4]{(b)}{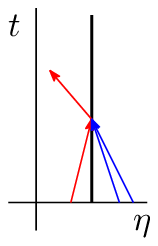} &
    \subfigimg[scale=1.4]{(c)}{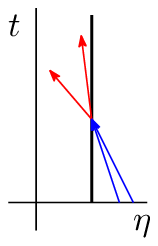}
  \end{tabular}
  \caption{Characteristics $\eta\pm(c_{1,2}\pm V)t=\text{const}$ of the
continuum problem in the moving frame with $\eta=x-Vt$ in phase 1 (blue) and phase 2 (red): (a) subkinks, $V<c_1$; (b) shocks, $c_1<V<c_2$;  (c)
superkinks, $V>c_2$. Here $\eta=x-Vt$.}
\label{fig:Characteristics}
\end{figure}

Fig.~\ref{fig:Characteristics} shows the arrangement of such characteristics in space-time for all three types of transition fronts. When $V<c_1$ (subkinks) or $V>c_2$ (superkinks), there are two incoming characteristics at the front, which reduces the number of unknowns to one, and therefore an additional condition is needed to find the remaining parameter, for instance, $V$. If $c_1<V<c_2$ (shocks), there are three incoming characteristics,  which means that all five parameters  can be determined without any additional conditions. In this sense kinks are undercompressive (non-Lax), while shocks are compressive \cite{lefloch2002hyperbolic}.
The necessity of an additional `kinetic relation' on discontinuous transition fronts was first pointed out in  \cite{truskinovskii1982equilibrium,truskinovskii1987dynamics,truskinovsky1993kinks}; see also \cite{abeyaratne2006evolution}.
The difference between subkinks and superkinks, which both require an additional condition closing the problem, is not apparent in this purely continuum setting.

\subsection{Dissipation rate}
While the system of  continuum  equations \eqref{eq:cont} is conservative, it known  that the corresponding discontinuous solutions may be dissipative. One way to supply the missing closure relations for subkinks and superkinks is to specify the dissipation rate at the moving transformation front.

For  all  three classes of fronts  the energy dissipation on the discontinuity can be written as a product \cite{truskinovskii1982equilibrium}:
\beq
{\cal R}=GV\geq 0,
\label{eq:R}
\eeq
where $V$ is the velocity of the front and  $G$ is the conjugate generalized (or driving) force, which is also known as the energy release rate. After appropriate symmetrization \cite{truskinovskii1987dynamics}, it takes the form
\begin{equation}
G=\llbracket {\cal \phi}(\varepsilon) \rrbracket-\{\sigma(\varepsilon)\}\llbracket \varepsilon\rrbracket,
\label{eq:G_def}
\end{equation}
where we introduced a notation for the averaging over the jump  $\{f\}=(f_{+}+f_-)/2$. The quasistatic notion of a driving force on a moving discontinuity dates back to Eshelby \cite{eshelby1970energy,knowles1979dissipation,heidug1985thermodynamics}.
A recent application of this notion in inertial dynamics can be found, e.g., in \cite{truskinovsky2008dynamics}.

In our piecewise linear continuum model the driving force $G$ can be computed explicitly. We obtain
\begin{equation}
G=\frac{E_2-E_1}{2}(\varepsilon_c^2-\varepsilon_{+}\varepsilon_{-})+\frac{\sigma_0}{2}(\varepsilon_{+}+\varepsilon_{-}-2\varepsilon_c).
\label{eq:Dissipation}
\end{equation}
In terms of the diagrams in Fig.~\ref{fig:stress_strain}, one can show that $G$ can be represented as the difference between the two colored areas between the Raleigh line and the stress-strain curve:  $G=S_2-S_1$.  Given that $V>0$, the  area  $S_1$ (blue) corresponds to the energy rate received on the jump  while  the area  $S_2$ (red) describes the rate of energy loss. To ensure the overall dissipative nature of the jump encapsulated by the inequality \eqref{eq:R}, it is therefore necessary  that $S_2 \geq S_1$.

Note that according to Fig.~\ref{fig:stress_strain}, in the case of subkinks the energy is received at the frontal part and lost (dissipated) at the back part of the transition front. Inside  shocks the energy can only  dissipated. For superkinks  the energy is lost in the frontal part and regained in the back part.

\subsection{Inner structure of the fronts}
As we have seen, in the continuum model the transition region is infinitely localized in space (jump discontinuity). However, the different arrangements   shown in Fig.~\ref{fig:stress_strain} suggest that it may be of interest to reconstruct the energetic structure of each of the archetypal front in the  configurational space of strains varying from $\varepsilon_{+}$ to $ \varepsilon_{-}$. The idea is that the energy transfers implied by the relative size of  the areas $S_1$ and  $S_2$ shown in Fig.~\ref{fig:stress_strain} are accomplished by some microscopic dispersive mechanisms that are overlooked by the continuum approximation.

For instance, in the case of subkinks, the continuously emerging energy in the frontal part of the transition region must be somehow transported from the back  of the front where it is released. Such transport can be accomplished  by the emitted sub-continuum (lattice) waves whose group velocity is larger than their phase velocity (which is necessarily equal to $V$). In the case of superkinks, the energy released in the frontal part is at least partially re-acquired  in the back part, and for this the system can use lattice waves whose group velocity is smaller than the phase velocity. At any rate, to support all the three types of the fronts, the dispersion must be sufficiently complex, which is of course the case for the original discrete model.
\begin{figure*}
  \centering
  \begin{tabular}{@{}p{0.3\linewidth}@{\quad}p{0.3\linewidth}@{\quad}p{0.3\linewidth}@{}}
    \subfigimg[width=\linewidth]{(a)}{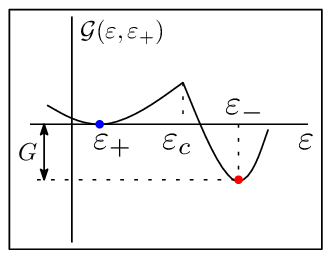} &
    \subfigimg[width=\linewidth]{(b)}{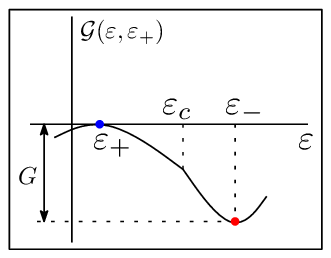} &
    \subfigimg[width=\linewidth]{(c)}{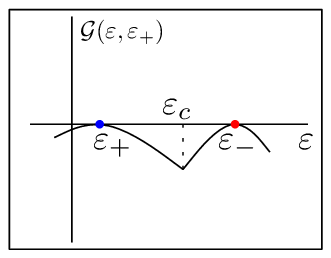}
  \end{tabular}
  \caption{Different behavior of the dissipation function $\mathcal{G}(\varepsilon,\varepsilon_{+})$: (a) subkinks, $V<c_1$, (b) shocks, $c_1<V<c_2$, (c) superkinks, $V>c_2$. }
\label{fig:psi}
\end{figure*}

To support this intuitive picture, it is instructive to introduce the notion of the local energy variation inside the strain interval connecting the limiting states  $\varepsilon_{+}$ and  $ \varepsilon_{-}$. Since the actual trajectory in the stress-strain space is not known, we can consider energy variation   along the Rayleigh line which ensures the conservation of the  macroscopic mass and momentum. The corresponding auxiliary function was introduced in \cite{truskinovsky2002nucleation} and in our notation it takes the form
\[
\mathcal{G}(\varepsilon, \varepsilon_{+} )=\phi(\varepsilon)-\phi(\varepsilon_{+})-(\varepsilon-\varepsilon_{+})\Sigma(\e),
\]
where
\[
\Sigma(\varepsilon)=\sigma(\varepsilon_{+})+\dfrac{\rho V^2}{2}(\varepsilon-\varepsilon_{+})
\]
is the average of $\sigma(\e_+)$ and the stress taken along the Rayleigh line. One can show that the limiting states $\varepsilon_{+}$ and $\varepsilon_{-}$ correspond to the extrema of the potential $\mathcal{G}$ with respect to $\e$. Note also that the reference energy is chosen in such a way that
\[
\mathcal{G}(\varepsilon_{+},\varepsilon_{+})=0,
\]
which means that the energy level assigned to the state ahead of the jump $\varepsilon=\varepsilon_{+}$ is zero. On the other hand, the overall dissipative (or non-dissipative) nature of each type of the fronts is reflected by the fact that at the final state $\varepsilon=\varepsilon_{-}$ we have
\[
\mathcal{G}(\varepsilon_{+},\varepsilon_{-})=-G \leq 0.
\]
In this way the implied energy landscape describes the energy variation inside the moving front independently of its type. However, it is important to  remember that the function $\mathcal{G}$ does not describe the actual variation of the energy inside the moving front, as we still do not refer to any particular  dispersive mechanisms operating inside the transition zone.

The behavior of  $\mathcal{G}$  as a function of $\varepsilon$ for all three types of transition fronts is shown schematically in
Fig.~\ref{fig:psi}.  As expected, the effective  energy landscapes for different universality classes are also qualitatively different.
Thus, for  subkinks, in addition to dissipation, which is expressed by the fact that the minimum at  $\varepsilon_{-}$ is lower than the minimum at $\varepsilon_{+}$, there is also an energy  barrier in between that needs to be overcome.  Crossing this barrier requires energy to be continuously transmitted by dispersion from the downstream, where it is continuously released. For shocks, there is no barrier, and the continuously released energy must be fully removed, with none of it being reabsorbed. Finally, for superkinks, there is no dissipation (as will be confirmed later). However, in this case there is an anti-barrier, and energy transmission by dispersion is still necessary, but now from upstream to downstream. Note also that since the barriers exist in the case of kinks and not shocks, the former can be considered as  topological `lattice defects',  while the latter remain non-topological.

\section{Quasicontinuum model}
\label{Quasicontinuum model}
The scale-free approximation we used to obtain the continuum model does not reveal the fate of the energy dissipated on the localized transition front and  does not explain which additional macroscopic jump condition must be chosen in the case of subkinks and superkinks.
To answer these questions we may simply solve the discrete problem.  The qualitative information can be also obtained from a quasicontinuum (QC) approximation  with sufficiently rich dispersion to adequately mimic the subcontinuum energy transport \cite{truskinovsky2006quasicontinuum,christov2007boussinesq}.

In this section we show that the minimal QC approximation of the FPU model capturing all of the dynamic regimes of interest can be constructed following the general approach  proposed in \cite{charlotte2012lattice}. The plan is to focus on temporal dispersion and introduce internal scales into the expression of   kinetic energy, while keeping the elastic energy as in the scale-free theory. The main idea dates back to the theory of rotational inertia of beams by Rayleigh \cite{rayleigh1877theory}, with subsequent generalizations for other dispersive problems  \cite{benjamin1972model,ostrovskii1977nonlinear}.
While in the context of discrete lattices, the QC theories of this type have been previously considered repeatedly \cite{collins1981quasicontinuum, rosenau1986dynamics,kevrekidis2002continuum,feng2004quasi}, we show below that even the simplest QC theory, targeting all three universality classes, must necessarily include some new elements.

\subsection{Main equations}
To construct  the QC approximation systematically, we  set $x=nh$, and introduce the variables $\varepsilon(x,t)=\varepsilon_n(t)$ and $\sigma(x,t)=\sigma(\varepsilon_n(t))$, viewed as functions of continuous space and time. We can then rewrite the infinite system \eqref{eq:EoM_strain} as a single advance-delay partial differential equation, which after the spatial Fourier transform takes the form
\beq
\rho h^2 \dfrac{d^2 \hat{\varepsilon}}{dt^2}=4\sin^2\left(\dfrac{kh}{2}\right)\hat{\sigma},
\label{eq:EoM_FT}
\eeq
where $\hat{f}(k,t)=\int_{-\infty}^\infty f(x,t)\exp{(ikx)}\,dx$ is the Fourier transform of $f(x,t)$. To simplify the problem and develop the corresponding long-wave asymptotic expansion, in what follows we assume that $kh \ll 1$.

To adequately describe the temporal dispersion \cite{charlotte2012lattice}, we use the Pad{\'e} expansion of $\sin^2(kh/2)$  in $kh$ which affects the kinetic energy, while preserving the classical continuum form of the elastic energy.  Keeping the terms in the denominator up to $O((kh)^4)$ yields
\begin{equation}
4\sin^2(kh/2)\approx \dfrac{(kh)^2}{1+a_1(kh)^2+a_2(kh)^4},
\label{eq:exp}
\end{equation}
where  $a_1= 1/12$ and $a_2= 1/240$. The need to retain exactly two subcontinuum terms in this expansion is dictated by the requirement that the resulting QC model is  minimal, as will be explained below. Substituting the expansion in Eq.~\eqref{eq:exp} into Eq.~\eqref{eq:EoM_FT} and mapping it back into physical space we obtain, after integration,
\begin{equation}
\rho \left(1-a_1\frac{\partial^2}{\partial x^2}+a_2\frac{\partial}{\partial x^4}\right)\frac{d^2 u}{dt^2} =\frac{\partial \sigma}{\partial x},
\label{eq:QCeqn_displ}
\end{equation}
where  $u(x,t)$ is the displacement  field  defined by the relation $u_x=\varepsilon$; here we also used the scaling $\tilde x=x/h$ but dropped the tildes in order to simplify the expressions. The single partial differential equation \eqref{eq:QCeqn_displ} represents the desired  QC approximation of the infinite FPU system \eqref{eq:EoM_displ} of ordinary differential equations.

To reveal the structure of the augmented kinetic energy term, we now derive the equation \eqref{eq:QCeqn_displ} from the Hamiltonian action principle.
We start with the sufficiently general action functional of the form
\begin{equation}
{\cal{A}} =\int_{\Omega} \mathcal{L}(u_{,i}\, , \, u_{,ij}\, , \, u_{,ijk})\, dq^1 dq^2,
\label{elementary_system1}
\end{equation}
where  $\mathcal{L}$ is a Lagrangian density, $q^1=x$ is the spatial coordinate, $q^2=t$ denotes time, and the subscripts after the comma indicate partial derivatives with respect to $q^1$ and $q^2$.   The integration in Eq.~\eqref{elementary_system1} is over the two-dimensional space-time domain $\Omega$ representing the evolving body between the time instants  $t=t_0$ and $t=t_1$. The deformation history is given by the function $u(q^a)$, $a=1,2$.
Given the structure of the action functional we can write the   Euler-Lagrange equations  in the form  \cite{gelfand63}:
\begin{equation}
\left(  \partial \mathcal{L}/\partial u_{,i} -  \left( \partial \mathcal{L}/\partial u_{,ij}\right)_{,j}+ \left( \partial \mathcal{L}/\partial u_{,ijk}\right)_{,jk}\right)_{,i}  = 0,
\label{elementary_system11}
\end{equation}
where here and in what follows the summation over repeated indices is implied. To obtain Eq.~ \eqref{eq:QCeqn_displ} from  Eq.~\eqref{elementary_system11},
we need to specify the Lagrangian. It is not difficult to see that the desired equation will be obtained if we consider
the Lagrangian in the form
 \begin{equation}
\mathcal{L}=  (\rho/2) (u_t^2 + a_1 u_{tx} ^2+a_2u_{txx} ^2)  -\phi(u_x).
\end{equation}
Here  the density of the elastic energy $\phi(u_x)$ is the same as in the classical continuum theory, while two sub-continuum terms with mixed derivatives appear in the expression of the kinetic energy. While the `micro-kinetic' term $a_1 u_{tx} ^2$ is now standard (see, e.g.,  \cite{theil2000study,kevrekidis2002continuum}), to our knowledge, the next term in the expansion, $a_2 u_{txx} ^2$, has not been used constructively before.

The advantage of using  the variational principle is that it allows one to derive not only the governing equation but also the corresponding jump conditions. This is relevant because despite regularization provided by the high derivative terms in the energy, our piecewise linear QC theory is still non-smooth at the transition point $u_x=\varepsilon_c$. The corresponding generalization of the RH jump conditions, compatible with our higher order  QC theory, emerges as a natural consequence of extremality of the action functional. Indeed, if the space-time domain $\Omega$ contains  a  surface $\Gamma$ of  discontinuity,  the standard Euler-Lagrange equations must  be supplemented by the additional necessary conditions of extremality localized on $\Gamma$. In our case the surface $\Gamma$ is characterized by the condition  $u_x=\varepsilon_c$, so  $\llbracket u \rrbracket=0$.  While the particle trajectories are continuous on $\Gamma$,  some derivatives of the displacement field may be discontinuous. We interpret the  constraints on such  singular surfaces  imposed by the action principle   as the dispersive Rankine-Hugoniot (DRH) jump conditions.

Using  the  standard manipulations detailed, for example, in \cite{gelfand63}, we obtain
\begin{equation}
\llbracket\partial \mathcal{L}/\partial u_{,i} -  \left( \partial \mathcal{L}/\partial u_{,ij}\right)_{,j}
+ \left( \partial \mathcal{L}/\partial u_{,ijk}\right)_{,jk} \rrbracket n_i  =0,
\label{elementary_system12}
\end{equation}
\begin{equation}
\llbracket\partial \mathcal{L}/\partial u_{,ij}- \left( \partial \mathcal{L}/\partial u_{,ijk}  \right)_{,k} \rrbracket n_i n_j=0.
\label{elementary_system122}
\end{equation}
\begin{equation}
\llbracket  \partial \mathcal{L}/\partial u_{,ijk}  \rrbracket n_i n_jn_k=0.
\label{elementary_system1222}
\end{equation}
Here  $n_a$ is the  unit vector normal to  $\Gamma$   facing the $+$ direction;  the spatial ($n_1$) and the temporal ($n_2$) components of such normal  are related through  $n_2=-n_{1}V$, where $V$ is the velocity of the discontinuity.

The necessary conditions \eqref{elementary_system12}, \eqref{elementary_system122} and \eqref{elementary_system1222} of extremality must be supplemented by the  kinematic compatibility conditions
\[
\llbracket  u_{,i} \rrbracket= \mu n_i,
\]
where  $ \mu$ is a scalar.  Eliminating $\mu$, we obtain an auxiliary jump relation
\begin{equation}
\llbracket u_t \rrbracket + V\llbracket u_x \rrbracket=0.
\label{eq:R-H_conditions0}
\end{equation}
which represents  the balance of mass across the discontinuity. In our special case
the three DRH conditions \eqref{elementary_system12}, \eqref{elementary_system122} and \eqref{elementary_system1222}  reduce to
\begin{equation}
 \rho V\llbracket u_t  -a_1 u_{txx} +a_2 u_{txxxx}\rrbracket +\llbracket \sigma(u_x)\rrbracket  =0,
\label{eq:R-H_conditions1}
\end{equation}
\begin{equation}
 \llbracket  a_1 u_{tx} -a_2 u_{txxx}\rrbracket =0
\label{eq:R-H_conditions2}
\end{equation}
\begin{equation}
 \llbracket   a_2 u_{txx}\rrbracket =0
\label{eq:R-H_conditions3}
\end{equation}
To satisfy all these conditions, we assume that $\llbracket u_t \rrbracket=0$ and $\llbracket u_{tx} \rrbracket=0$. Then  $\llbracket   u_{x} \rrbracket =0$, while the  two conditions \eqref{eq:R-H_conditions2} and \eqref{eq:R-H_conditions3} reduce to  $\llbracket  u_{txxx}\rrbracket =0$ and  $ \llbracket u_{txx}\rrbracket =0$, respectively.
The condition \eqref{eq:R-H_conditions1} reduces to
\begin{equation}
 \rho a_2V\llbracket  u_{txxxx}\rrbracket +\llbracket \sigma(u_x)\rrbracket  =0.
\label{eq:R-H_conditions11}
\end{equation}
The derived jump conditions guarantee that the physical description of the phenomena  in  the bulk  and  on the discontinuity surface are exactly the same.

\subsection{Dimensionless formulation}
In what follows,  we use dimensionless variables
\[
\tilde{V}= \dfrac{V}{c_1}, \quad \tilde{\sigma}= \dfrac{\sigma}{E_1}, \quad \tilde{\sigma}_0=\dfrac{\sigma_0}{E_1},
\]
with  tildes dropped to simplify notation. The system is controlled by the dimensionless parameters $\e_c$ and
\[
\gamma=\sqrt{\dfrac{E_2}{E_1}}>1.
\]
 For the analysis presented below, it is convenient to work with the following equation obtained by differentiating the dimensionless version of Eq.~\eqref{eq:QCeqn_displ} with respect to $x$:
\begin{equation}
\left(1-a_1\frac{\partial^2}{\partial x^2}+a_2\frac{\partial}{\partial x^4}\right)\frac{d^2 \varepsilon}{dt^2} =\frac{\partial^2 \sigma}{\partial x^2}
\label{eq:QCeqn_strain}
\end{equation}

\subsection{Traveling waves}
To find steadily moving transition fronts, we seek solutions of Eq.~\eqref{eq:QCeqn_strain} in the form of traveling waves:
\begin{equation}
\e(x,t)=\e(\eta), \quad \eta=x-Vt.
\label{eq:TWansatz}
\end{equation}
We place the front separating two linear regimes at $\eta=0$ and thus require that the following consistency condition is satisfied:
\beq
\varepsilon(0)=\varepsilon_c.
\label{eq:consistency}
\eeq
Moreover, we consider the solutions admissible  only if they satisfy  the inequalities
\beq
\varepsilon(\eta)>\varepsilon_c \;\; \text{for $ \eta<0$}, \quad  \varepsilon(\eta)<\varepsilon_c \;\; \text{for $\eta>0$}.
\label{eq:admissibility}
\eeq
Since our solutions can be expected to contain phonon radiation at $\pm \infty$, we formulate the boundary conditions in the form
\beq
\langle\varepsilon(\eta)\rangle\to\varepsilon_{\pm} \quad \text{as $\eta\to\pm\infty$},
\label{eq:BCs}
\eeq
with constant limits $\varepsilon_\pm$ constrained by the standard RH condition \eqref{eq:RH} with stress-strain law given by Eq.~\eqref{eq:stress-strain}, which in the dimensionless formulation becomes
\beq
\e_{-}=\dfrac{(V^2-1)\e_{+}-\sigma_0}{V^2-\gamma^2}.
\label{eq:RH_PL}
\eeq
The angular brackets in Eq.~\eqref{eq:BCs}
denote the average over the period of the short-wave oscillations representing phonon radiation.
The admissibility condition \eqref{eq:BCs} requires that $\varepsilon_{+}<\varepsilon_c$, and $\varepsilon_{-}>\varepsilon_c$. Physically, this  means that the moving transition front performs the switching from one branch of the piecewise linear stress-strain curve to another.

Substituting Eq.~\eqref{eq:TWansatz} into Eq.~\eqref{eq:QCeqn_strain}, integrating   twice and taking into account the boundary conditions \eqref{eq:BCs}, we obtain the odinary differential equation
\begin{equation}
V^2\left[1-a_1\frac{d^2}{d\eta^2}+a_2\frac{d^4}{d\eta^4}\right]\varepsilon(\eta)=\sigma(\eta)+(V^2-1)\varepsilon_{+},
\label{QCeqn2}
\end{equation}
where
\beq
\sigma(\eta)=\varepsilon(\eta)H(\eta)+(\gamma^2 \varepsilon(\eta)-\sigma_0)H(-\eta),
\label{eq:sigma}
\eeq
and $H(\eta)$ is the Heaviside function. We also need to apply the following jump conditions at $\eta=0$:
\beq
\llbracket \varepsilon\rrbracket=\left\llbracket  d\varepsilon/d\eta\right\rrbracket=0,
\label{eq:cont_conds}
\eeq
\beq
 \left\llbracket  d^3\varepsilon/d\eta^3\right\rrbracket=0, \quad
\left\llbracket  d^2\varepsilon/d\eta^2\right\rrbracket=0.
\label{eq:DRH_conds}
\eeq
It is straightforward to check that the condition \eqref{eq:R-H_conditions11}, which takes the form $\left\llbracket \sigma(\eta)\right\rrbracket-a_2 V^2\left\llbracket  d^4\varepsilon/d\eta^4\right\rrbracket=0$,  is satisfied automatically.
\begin{figure*}
 \begin{tabular}{@{}p{0.3\linewidth}@{\quad}p{0.3\linewidth}@{\quad}p{0.3\linewidth}@{}}
    \subfigimg[width=\linewidth]{(a)}{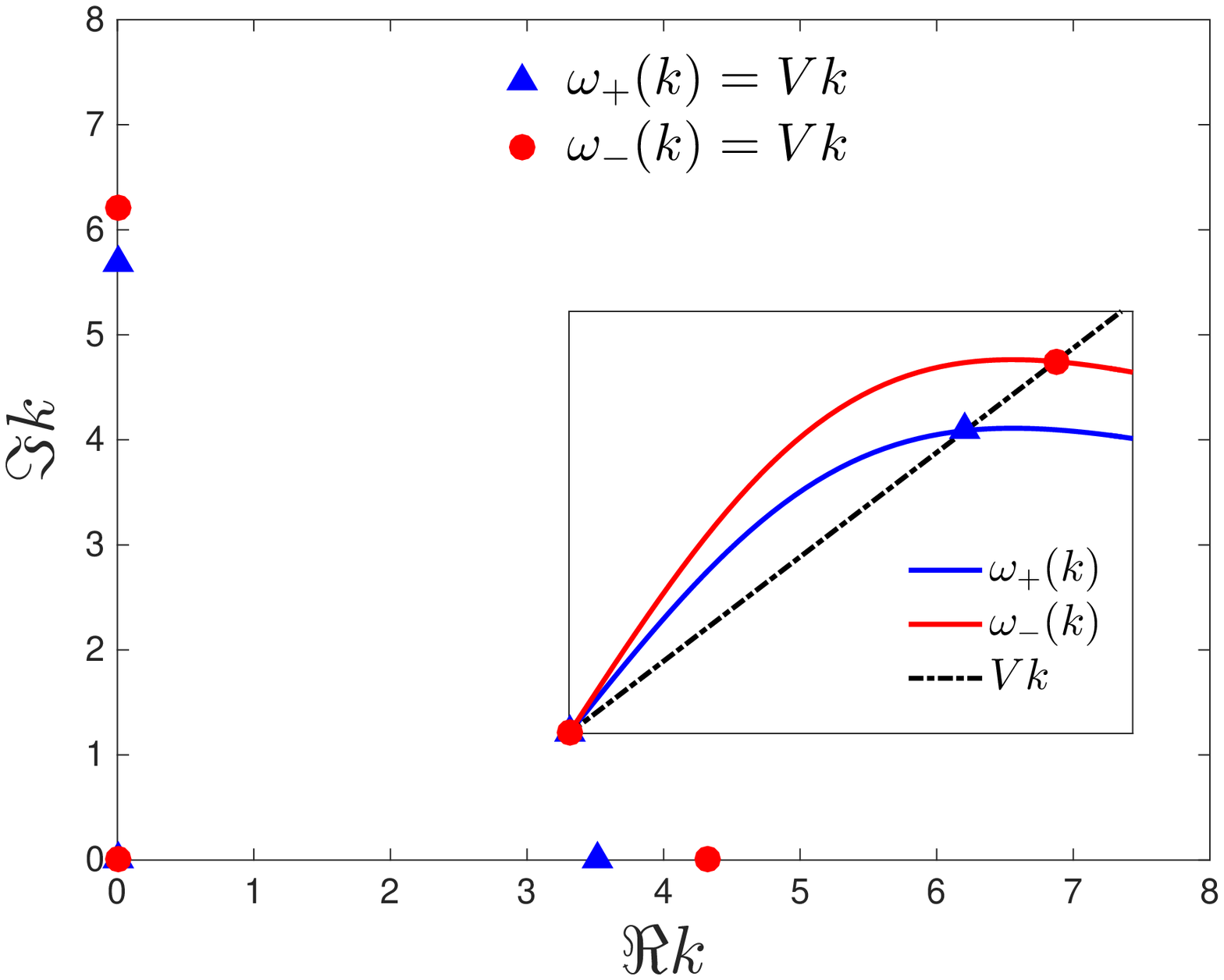} &
    \subfigimg[width=\linewidth]{(b)}{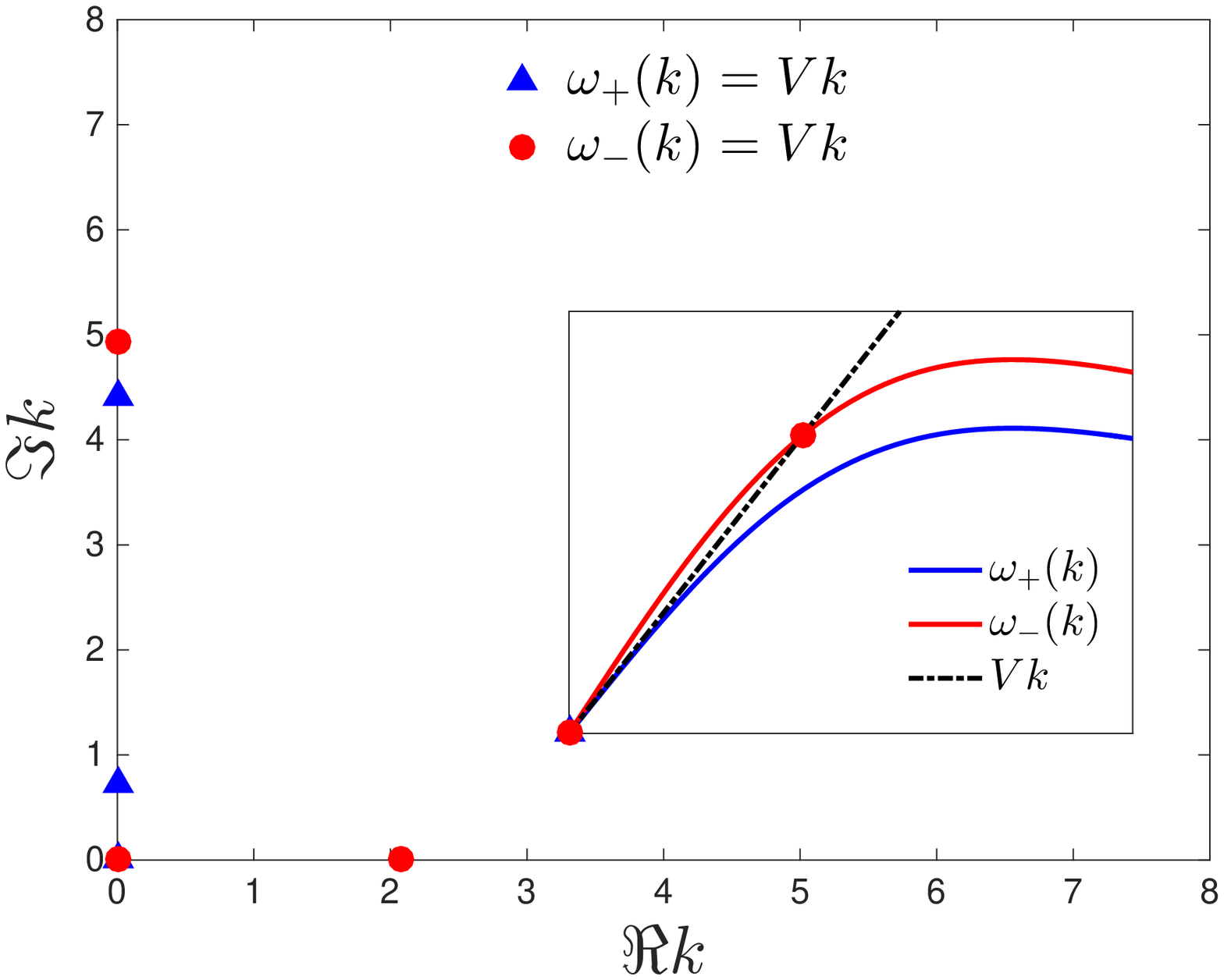} &
    \subfigimg[width=\linewidth]{(c)}{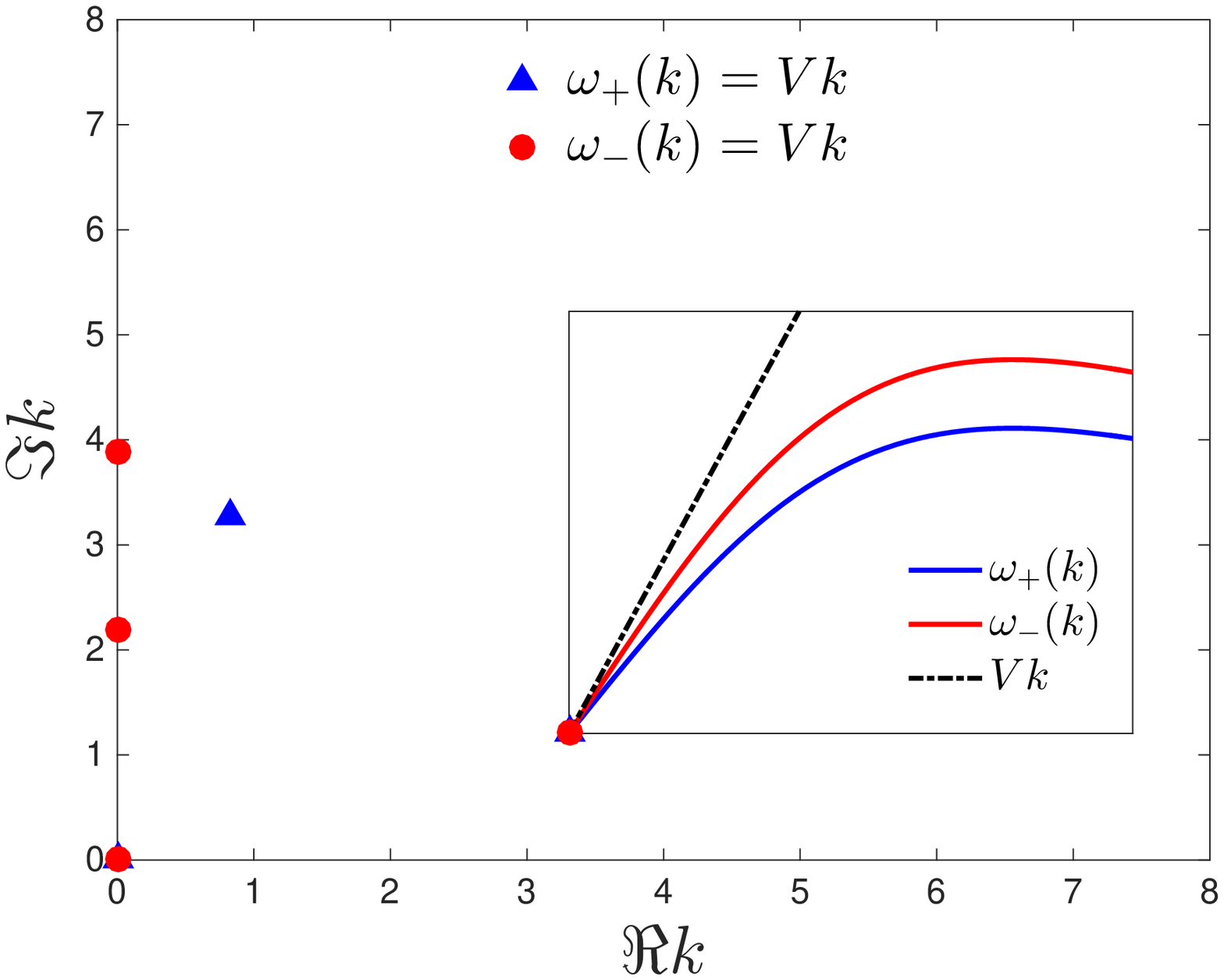}
  \end{tabular}
  \caption{The  characteristic roots in the quasicontinuum model when (a) $V<1$, (b) $1<V<\gamma$, (c) $V>\gamma$. Due to symmetry, only the roots with $\Im k \geq 0$ and $\Re k \geq 0$ are shown. Insets show the dispersion relations and real roots as intersections with the line $Vk$.}
\label{fig:Roots_quasi}
\end{figure*}

\subsection{Mechanical radiation}
Since Eq.~\eqref{QCeqn2} is piecewise linear, it can be solved explicitly. The analytical solution in each of the two linear regimes can be written as a combination of linear waves whose frequencies and wave numbers satisfy the characteristic equations
\beq
\omega^2_{\pm}(k)-V^2k^2=0,
\label{eq:roots}
\eeq
where $\omega_{+}(k)$ and $\omega_{-}(k)$  are the dispersion relations defined by
\beq
\dfrac{\omega_+^2(k)}{k^2}=\dfrac{\omega_-^2(k)}{(\gamma k)^2}=\dfrac{1}{1+a_1k^2+a_2k^4}
\label{eq:dispersion_QC}
\eeq
and shown in the insets of Fig.~\ref{fig:Roots_quasi}. The double root of \eqref{eq:roots} at $k=0$ is responsible for a linear term in the solution, and due to the assumption of  boundedness,  it contributes only constants in each domain of linearity. Due to the even symmetry of the functions $\omega_{\pm}(k)$,
the four nonzero roots of \eqref{eq:roots}, which we denote by $k_j^{\pm}$, $j=1,2,3,4$, must satisfy $k^{\pm}_3=-k^{\pm}_1$ and $k^{\pm}_4=-k^{\pm}_2$. Therefore it suffices to seek nonzero roots with $\Im k>0$ and $\Re k>0$, where $\Re k$ and  $\Im k$ are real and imaginary parts of $k$, respectively. The structure of the roots for three different types of fronts is shown in Fig.~\ref{fig:Roots_quasi}.

Of principal importance for the description of phonon radiation produced by the moving front are the nonzero real roots of \eqref{eq:roots}. The corresponding points of intersection of $\omega_{\pm}(k)$ and $Vk$ are marked in the insets of Fig.~\ref{fig:Roots_quasi}. When $V<1$ (subkinks), a symmetric pair of such roots $\pm k^{\pm}$  exists for each domain of linearity: when $1<V<\gamma$ (shocks), only the roots $\pm k^{-}$ remain, and, finally in the case of superkinks $V>\gamma$ there are no nonzero real roots at all. Since each nonzero real root describes  energy radiation to and from the moving front, the superkinks can potentially receive but cannot dissipate energy in the form of radiated waves.

To exclude the energy flux from infinity (anti-dissipation, which can be sometimes interpreted  as  an AC driving \cite{gorbushin2020frictionless}), we must   impose the  radiation  conditions disqualifying some of the waves associated with the real  roots. In our case these conditions,  comparing the velocity of the energy propagation (group velocity)  with the velocity of the front, take the   form \cite{slepyan2012models,truskinovsky2005kinetics}
\beq
\omega_+'(k)>V, \quad \omega_-'(k)<V.
\label{eq:rad_conds}
\eeq
Since the functions $\omega_{\pm}(k)$ are known, these conditions are explicit. They leave only one real root component of the solution in the case of subkinks and shocks.

\subsection{General solution}
We observe that the whole  configuration of the roots of the characteristic equations (real and complex) changes depending on the values of $V$. The roots with $\Re k>0$ and $\Im k>0$ are given by
\beq
\begin{split}
&k_{1,2}^+=\sqrt{2}\sqrt{-5\mp \dfrac{\sqrt{5(7V^2-12)}}{V}}, \\
&k_{1,2}^-=\sqrt{2}\sqrt{-5\mp \dfrac{\sqrt{5(7V^2-12\gamma^2)}}{V}}
\end{split}
\label{eq:QCroots}
\eeq
and correspond to real, purely imaginary or complex pairs as $V$ varies. More specifically, for the state ahead  of the moving front we have the following three regimes:
\begin{equation}
\begin{gathered}
k^+_1=ip,\quad k^+_2=s,\quad V<1,\\
k^+_{1,2}=ip_{1,2},\quad 1<V<V_{*},\\
k^{+}_{1,2}=id\pm f,\quad V>V_{*}
\end{gathered}
\label{eq:roots_omega_1}
\end{equation}
For the state behind the front  we have the same three regimes but in different $V$ ranges:
\begin{equation}
\begin{gathered}
k^{-}_1=iq,\quad k_2^{-}=r,\quad V<\gamma,\\
k^{-}_{1,2}=iq_{1,2},\quad \gamma<V<V_{**},\\
k^{-}_{1,2}=ig\pm w,\quad V>V_{**},
\end{gathered}
\label{eq:roots_omega_2}
\end{equation}
 Explicit expressions for the real and positive functions  $p(V)$, $s(V)$, $p_{1,2}(V)$, $d(V)$, $f(V)$, $q(V)$, $r(V)$, $q_{1,2}(V)$, $g(V)$ and $w(V)$ can be extracted from \eqref{eq:QCroots}.  The critical values
\[
V_*=\sqrt{ 12/7}, \quad V_{**}=\gamma \sqrt{ 12/7}>V_*,
\]
are the artifacts of the QC approximation, which, as we show below, do not have any fundamental meaning.

Applying the radiation conditions \eqref{eq:rad_conds} and the boundary conditions \eqref{eq:BCs}, we can  write the general solutions  corresponding to all three types of transition fronts. In particular, in the case of subkinks ($V<1$), the solution takes the form
\begin{equation}
\begin{split}
&\varepsilon(\eta)=\\
&\begin{cases} \varepsilon_{-}+B_1e^{q\eta}+B_2\cos(r\eta)+B_3\sin(r\eta), & \eta<0\\
              \varepsilon_{+}+A_1e^{-p\eta}, & \eta>0.
\end{cases}
\end{split}
\label{eq:QCsoln_subkinks}
\end{equation}
One can see that  for subkinks  there is one unknown coefficient on the $+$ side and three on the $-$ side. All
of them can be found from the consistency, continuity, RH and DRH conditions. Indeed, the consistency condition \eqref{eq:consistency} and the first of the continuity conditions in Eq.~\eqref{eq:cont_conds} yield in this case
the relations
\beq
\varepsilon_{+}+A_1=\varepsilon_c=\varepsilon_{-}+B_1+B_2.
\label{eq:cons_subkinks}
\eeq
This allows us to eliminate $\e_{\pm}$. Using the RH condition \eqref{eq:RH_PL}, the second continuity condition in Eq.~\eqref{eq:cont_conds} and the DRH conditions \eqref{eq:DRH_conds}, we then obtain the system of  linear equations for the coefficients in Eq.~\eqref{eq:QCsoln_subkinks}:
\[
\begin{gathered}
-C_0 A_1+B_1+B_2=b\\
pA_1+qB_1+rB_3=0,\\
p^2A_1-q^2B_1+r^2B_2=0,\\
p^3 A_1+q^3 B_1-r^3 B_3=0,
\end{gathered}
\]
where
\[
C_0=\frac{V^2-1}{V^2-\gamma^2},\quad  b= (1-C_0)\varepsilon_c+\frac{\sigma_0}{V^2-\gamma^2}.
\]
The system yields explicit expressions for the four unknown coefficients $A_1$, $B_1$, $B_2$ and $B_3$ as functions of $V$ that are not provided here to simplify the exposition. The expressions for $\varepsilon_{\pm}(V)$ are then found from Eq.~\eqref{eq:cons_subkinks}.

For shocks and superkinks ($V>1$) the structure of the roots in Eq.~\eqref{eq:roots_omega_1} and Eq.~\eqref{eq:roots_omega_2} changes depending on the value of $V$ relative to the thresholds $V_*$ and $V_{**}$. To account for this, it is convenient to introduce the shortcuts
\[
\lambda_{1,2}=\begin{cases}-p_{1,2}, & 1<V<V_*\\
-d\pm if, & V>V_*
\end{cases}
\]
and
\[
\mu_{1,2}=\begin{cases} q_{1,2}, & \gamma<V<V_{**}\\
g\pm iw, & V>V_{**}.
\end{cases}
\]
Then for shocks ($1<V<\gamma$) we have
\begin{equation}
\begin{split}
&\varepsilon(\eta)=\\
&\begin{cases} \varepsilon_{-}+B_1e^{q\eta}+B_2\cos(r\eta)+B_3\sin(r\eta), & \eta<0 \\
\varepsilon_{+}+A_1e^{\lambda_1\eta}+A_2e^{\lambda_2\eta}, & \eta>0,
\end{cases}
\end{split}
\label{eq:QCsoln_shocks}
\end{equation}
with two unknown coefficients on $+$ side and three
on $-$ side. The conditions \eqref{eq:consistency}, \eqref{eq:RH_PL}, \eqref{eq:cont_conds} and \eqref{eq:DRH_conds} yield
$\varepsilon_{+}+A_1+A_2=\varepsilon_c=\varepsilon_{-}+B_1+B_2$ and
the following linear system for the coefficients in Eq.~\eqref{eq:QCsoln_shocks}:
\begin{equation}
\begin{gathered}
-C_0(A_1+A_2)+B_1+B_2=b\\
\lambda_1A_1+\lambda_2A_2-qB_1-rB_3=0,\\
\lambda_1^2A_1+\lambda_2^2A_2-q^2B_1+r^2 B_2=0,\\
\lambda_1^3 A_1+\lambda_2^3 A_2-q^3 B_1+r^3 B_3=0.
\end{gathered}
\label{eq:shock_coefs}
\end{equation}
This system of four equations does not allow one to find all five unknown
coefficients $A_1$, $A_2$, $B_1$, $B_2$ and $B_3$ as functions of $V$. In other words, the structure of shocks is not fully determined internally,
which in turn means that $\e_{\pm}$ cannot be determined as functions of $V$. All parameters are fully defined in this case only if we provide one additional external condition, for example, $\varepsilon_{+} = 0$, which implies  $A_1+A_2 = \varepsilon_c$.

In the range $V>\gamma$ (superkinks) the solution reads
\begin{equation}
\varepsilon(\eta)=\begin{cases}\varepsilon_{-}+B_1e^{\mu_1 \eta}+B_2e^{\mu_2 \eta}, & \eta<0\\
\varepsilon_{+}+A_1e^{\lambda_1 \eta}+A_2e^{\lambda_2 \eta}, & \eta>0.
\end{cases}
\label{eq:QCsoln_superkinks}
\end{equation}
In this case there are two unknown coefficients on each side of the front, so the solution is again fully specified by conditions
\eqref{eq:consistency}, \eqref{eq:RH_PL}, \eqref{eq:cont_conds} and \eqref{eq:DRH_conds}, which yield the linear system
\[
\begin{gathered}
-C_0(A_1+A_2)+B_1+B_2=b,\\
\lambda_1A_1+\lambda_2A_2-\mu_1B_1-\mu_2B_2=0,\\
\lambda_1^2A_1+\lambda_2^2A_2-\mu_1^2B_1-\mu_2^2B_2=0,\\
\lambda_1^3A_1+\lambda_2^3A_2-\mu_1^3B_1-\mu_2^3B_2=0.
\end{gathered}
\]
for the four unknown coefficients $A_1$,$A_2$, $B_1$, $B_2$ that can be found as explicit functions of $V$, as well as the relations $\varepsilon_{+}+A_1+A_2=\varepsilon_c= \varepsilon_{-}+B_1+B_2$, which allows one to find the two remaining functions $\varepsilon_{\pm} (V)$.

To summarize, after using the conditions \eqref{eq:rad_conds}, \eqref{eq:BCs}, \eqref{eq:consistency} and the first condition in Eq.~\eqref{eq:cont_conds}, we are left in the range $V<1$ (subkinks) with one unknown coefficient on $+$ side and three on $-$ side (a single exponential boundary layer and a radiated wave). All of them can be found from the four conditions: the second condition in Eq.~\eqref{eq:cont_conds}, Eq.~\eqref{eq:RH_PL} and Eq.~\eqref{eq:DRH_conds}. When $1<V<\gamma$ (shocks) we are left with two coefficients on $+$ side and three on $-$ side (a radiated wave and a single exponential boundary layer)
and only four conditions. This leaves one of the constants in the corresponding linear system \eqref{eq:shock_coefs} undetermined. Finally, in the range  $V>\gamma$ (superkinks) there are two coefficients on each side, so the solution is again fully specified by the four conditions.

Once the strain field is determined in each regime, particle velocity is found from $v(\eta)=-V\e(\eta)$.

\subsection{Discussion}
Now that the mathematical structure of traveling wave solutions is well understood, we provide a physical interpretation of the results that furnishes  a somewhat more intuitive explanation of the fundamental differences between the three types of transition fronts.

Observe first that in all three cases, the traveling wave solutions describing the transition fronts can be written in the same general form
\begin{equation}
\varepsilon(\eta)=
\varepsilon_{\pm}+\Lambda_{\pm}(\eta)+\Phi_{\pm}(\eta),\quad \eta\gtrless
0,
\label{eq:soln_form}
\end{equation}
Here the functions $\Lambda_{\pm}(\eta)$  depend on the real roots of the characteristic equation and describe the radiative part of the solution. The functions $\Phi_{\pm}(\eta)$ depend on the non-real complex roots and describe the exponentially localized boundary layers on both sides of the moving fronts. The constant terms in \eqref{eq:soln_form} are due to the double root at the origin; the  strains  $\varepsilon_{\pm}$  correspond to the averaged states at $\eta \to \pm \infty$ and satisfy the classical RH condition \eqref{eq:RH_PL}.

We now consider in more detail the radiative component of the solution $\Lambda_{\pm}(\eta)$. We have seen that to exclude the energy flux from infinity (radiation condition), we  need to set (in all three cases) that radiation is absent ahead of the front, so that
\[
\Lambda_+ (\eta)=0.
\]
Moreover, while all three solutions obtained above in equations \eqref{eq:QCsoln_subkinks}, \eqref{eq:QCsoln_shocks} and \eqref{eq:QCsoln_superkinks}, have the form \eqref{eq:soln_form}, the nontrivial radiation component (behind the moving front) exists only for subkinks and shocks and can be written as
\begin{equation}
\Lambda_{-}(\eta)=2\alpha^{-}\cos{(r\eta+\beta^{-})},
\label{eq:radiation}
\end{equation}
with $\alpha^{-}$, $\beta^{-}$ expressed in terms of $B_2$ and $B_3$ in Eq.~\eqref{eq:QCsoln_subkinks} and Eq.~\eqref{eq:QCsoln_shocks}. Thus,
both subkinks and shocks radiate (dissipate) energy.
In contrast, the superkinks are completely free from radiation (dissipation), since in this case we also have $\Lambda_{-}(\eta)=0$.

We now turn to the boundary layer terms $\Phi_{\pm}(\eta)$. For subkinks it involves a single decaying exponential term on each side of the front ($\Phi_{+}(\eta)=A_1e^{-p\eta}$, $\Phi_{-}(\eta)=B_1e^{q\eta}$); see Eq.~\eqref{eq:QCsoln_subkinks}.  For shocks, there is a single exponential decay behind the front ($\Phi_{-}(\eta)=B_1e^{q\eta}$), while ahead of it the decay is double exponential ($\Phi_{+}(\eta)=A_1e^{p_1\eta}+A_2e^{p_2\eta}$) when $1<V<V_*$ and oscillatory ($\Phi_{+}(\eta)=e^{-d\eta}[A_1\cos(f\eta)+A_2\sin(f\eta)]$) when $V_*<V<\gamma$ (see Eq.~\eqref{eq:QCsoln_shocks}). For superkinks, there is a similar transition from double exponential to oscillatory decay ahead of the front at $V=V_{*}$ if $\gamma<V_*$ and behind it at $V=V_{**}$ (see Eq.~\eqref{eq:QCsoln_superkinks}). As the analysis of the discrete problem presented below shows, both double exponential and oscillatory decay are artifacts of the chosen QC approximation.

As we have seen, for both types of kinks all parameters of the traveling wave, and in particular, the limiting states $\varepsilon_{\pm}$, are fully determined by the front velocity $V$. This means that the kinetic relations $G=G(V)$, whose absence in the classical continuum description produced the fundamental ill-posedness of the problem, are now fixed through the recovery of the internal structure of the kinks. In other words, such fronts are fully  autonomous in the sense that their kinetics is fully controlled by the microscopic  dispersion. For instance, if the state in front of the moving kink $\varepsilon_{+}$ is known, then both the state behind, $\varepsilon_{-}$, and the velocity of the front $V$ are fully determined.

In contrast, in the case of  shocks, the knowledge of $V$ is not sufficient to determine both $\varepsilon_{\pm}$, and one of the
limiting strains remains as a free parameter. As a result, no particular kinetic relation in the form  $G=G(V)$ emerges from the reconstruction of the internal structure of such transition front. In other words, in the case of shocks, the knowledge of the state ahead is not sufficient for complete specification of the remaining  parameters and for fixing the internal structure of the transition. This means, for instance,  that in  addition  to the state ahead of the front $\varepsilon_{+}$, another piece of information has to be prescribed by the external (non-traveling-wave) solution in order to make the front velocity $V$ known.

\subsection{Characteristics}
The obtained QC picture is in full agreement with what we have learned by studying the classical continuum approximation in Sec.~\ref{Continuum model}. There we found that kinks are different from shocks primarily due to the difference in the number of incoming characteristics shown in Fig.~\ref{fig:Characteristics}.

In particular, Fig.~\ref{fig:Characteristics} shows that for both types of kinks two characteristics are bringing information to the front. Since in our analysis of the internal structure of the transition fronts we eliminated particle velocities $v(\eta)$, we may always assume  that this information concerns the limiting values $v_{\pm}$. Therefore, we can conclude that in the case of kinks, no information about one of the limiting strains $\varepsilon_{\pm}$ is arriving from outside. Thus, to fix the unknown limiting strain and to ultimately specify the front velocity $V$,  the system must rely exclusively on the internal dispersive machinery. The analysis of the QC approximation shows that such machinery is indeed in place delivering all of the unknown quantities.

In contrast, in the case of shocks, the classical continuum model tells us that the three characteristics are coming from outside. Therefore, the system can use one additional piece of external information to fix the limiting strains   $\varepsilon_{\pm}$  and to specify the front velocity $V$. In this case, the
internal dispersive structure of the front does not have an autonomy and simply adjusts to the conditions imposed from the outside. Remarkably, this is exactly what our study of the dispersive QC model have shown: for shocks the internal traveling wave  solution is (one-parameter) underdetermined, and to make the global problem well posed a single additional piece of information is needed. Such information is then naturally  provided by the additional incoming characteristic which does exists in the case of shocks.

\subsection{Dynamical system}
Since all three types of transition fronts represent traveling wave solutions of the fourth order ordinary differential equation \eqref{QCeqn2}, it is of interest to examine them from the point of view of the theory of dynamical systems. In this perspective they emerge as fundamentally different types of heteroclinic trajectories  connecting various types of attractors in the four-dimensional phase space. The nature of such attractors depends on the structure of the roots of the characteristic equations, which control the asymptotic behavior of the  heteroclinic trajectories as $\eta \to \pm \infty$. The knowledge of these asymptotics is sufficient to  distinguish between the different universality classes of the  transition fronts.

For example, in the case $V<1$ (subkinks) the transition fronts correspond to heteroclinic trajectories of the type center-saddle to center-saddle. Such transitions are non-generic and are possible due to the sufficiently high dimensionality of our dynamical system.  More specifically, they  are captured by our QC approximation because the latter includes  the minimal number of the higher order dispersive corrections to the classical continuum model    which makes the corresponding phase space four-dimensional. At $\eta=-\infty$ the heteroclinic trajectory describing subkinks unwinds as the center-related separatrix.  The corresponding two-dimensional center effectively describes the radiation behind the moving subkink, while the saddle-related component of the asymptotics describes the exponent boundary layer. At $\eta=+\infty$ this trajectory ends as a saddle-related separatrix, which describes the exponential boundary layer ahead of the moving front.

Similar considerations can be applied to shock and superkink trajectories. For simplicity, we assume in what follows that $\gamma<\sqrt{12/7}$ and $V<\sqrt{12/7}$. This eliminates the oscillatory decay for shocks and superkinks, which, as we have discussed, is an artifact of the QC approximation.
In the range $1<V<\gamma$ (shocks) the corresponding heteroclinic orbits are  of the type center-saddle to saddle-saddle. Such transitions are clearly generic. At $\eta=-\infty$  the  heteroclinic trajectory  unwinds again as a center-related separatrix describing radiation behind the front.  The center-related part of the asymptotics describes again the exponential boundary layer.  At $\eta = + \infty$ the trajectory  ends as a saddle related separatrix describing the exponential decay ahead of the front. Finally, for $V>\gamma$ (superkinks) the corresponding orbit is of saddle-saddle to saddle-saddle type. Such transitions are again non-generic. In this case the heteroclinic trajectory starts as a saddle-related separatrix  describing the exponential decay behind the front and ends again as a saddle-related separatrix  describing the boundary layer ahead of the front.

We have thus confirmed that the physical nature of the all three types of the transition fronts described by the general Eq.~\eqref{eq:QCeqn_displ} is fully consistent with  the asymptotic behavior of the heteroclinic trajectories  at  $\eta \to \pm \infty$.  The fact that the latter is controlled by the structure of the roots of the characteristic equations characterizing  the corresponding attractors goes beyond the adopted piecewise linear approximation of the stress-strain relation.   Thus, even without such an assumption the subkinks can be expected to correspond to non-generic transition fronts, which are  described by center-saddle to center-saddle trajectories and which generate their own kinetic relations.  Such transitions, however, would be possible only if sufficiently  higher order  dispersion is included into the model. Similarly, even in a smoother model shocks correspond (under our assumptions) to the heteroclinic orbits that  are generic saddle-saddle to center-saddle trajectories, which do not generate any specific kinetic relations. Finally, under the same assumptions superkinks  are non-generic transitions described by the saddle-saddle to saddle-saddle heteroclinic orbits. The fact that all possible  types of sufficiently low-dimensional non-dissipative attractors are accounted for, suggests that the proposed classification of the transition fronts is exhaustive.

\subsection{Dissipation rate}
In the dispersively regularized setting the jump discontinuities of strain and velocity that are present in the classical continuum theory are replaced by the extended transition zones. In addition, the energy released on such jumps in the continuum theory no longer disappears locally. Instead it is channeled by nonlinearity from long to short waves and radiated away from the moving front in the form of lattice waves. In the piecewise linear theory it is transported by such waves to infinity. In other words, despite the absence of explicit damping, the effective dissipation takes place due to the energy escape by phonon radiation.

The developed QC model allows one to trace all these processes in full detail. In particular, one can compute explicitly the thermodynamic driving force $G$ for all three types of transition fronts and determine the corresponding rate of energy dissipation ${\cal R}=GV \geq 0$.  Based on the analysis of the existing modes of radiation, one can see that ${\cal R}$ is strictly positive for subkinks and shocks but equals zero for superkinks.

More specifically, depending on the structure of the real roots of the characteristic equations, the transition front may or may not emit elastic waves. In general, we have ${\cal R}={\cal R}_+ +{\cal R}_-$, where
\beq
\begin{gathered}
{\cal R}_+ = \sum_{k \in {\cal N}_+}\langle {\cal E}_{+}(k)\rangle(\omega_+'(k)-V)=G_{+}V,\\
{\cal R}_- = \sum_{k \in {\cal N}_-}\langle {\cal E}_{-}(k)\rangle(V-\omega_-'(k))=G_{-}V,
\end{gathered}
\label{eq:Rpm_gen}
\eeq
and $G_+$ and $G_-$ are the cumulative energy fluxes associated with emitted elastic waves ahead and behind the front, respectively.
Here ${\cal N}_{\pm}=\{k:\,\Im k=0,\,\Re k>0,\, \omega_{\pm}(k)=Vk,\,\omega'_{\pm}(k)\gtrless V\}$ is the set of positive real roots of the characteristic equation for corresponding linear regime that satisfy the radiation conditions \eqref{eq:rad_conds}, and ${\cal E}_{\pm}(k)$ are the energy densities associated with the corresponding modes, averaged over the corresponding time period $T=2\pi/\omega_{\pm}(k)$, with $\langle f \rangle=T^{-1}\int_0^T f dt$.
The energy is transported away from the front with relative velocities $\omega_{\pm}'(k)-V$ \cite{brillouin1953wave}.

From the structure of the exact solutions of the QC model one can see that the set ${\cal N}_+$ is empty for all transition fronts. Thus, independently of the front type there is no radiation of phonons ahead of the front, and $G_+=0$. In the superkink regime, ${\cal N}_{-}$ is also empty, and therefore $G_{-}=0$ as well, yielding ${\cal R}=0$. We recall that in the case of subkinks and shocks there is a single emitted lattice wave mode with wave number $r>0$ propagating in the region $\eta<0$, so that ${\cal N}_{-}=\{r\}$. The associated energy with the density
\[
{\cal E}_{-}(r)=\dfrac{V^2}{2}\left(\Lambda_{-}^2+a_1(\Lambda_{-}')^2+a_2(\Lambda_{-}'')^2\right)+\dfrac{\gamma^2}{2}\Lambda_{-}^2.
\]
averaged over the period $2\pi/\omega_{-}(r)$, is transported backwards relative to the moving front with the relative velocity $\omega_{-}'(r)-V$ \cite{brillouin1953wave}. This yields the driving force $G=G_{-}+G_{+}$ given by
\[
G=G_{-}=2\gamma^2(\alpha^-)^2\omega_{-}^2(r)\left(1-\dfrac{\omega_{-}'(r)}{V}\right)> 0,
\]
where we recall that $\alpha^-$ is half of the amplitude of the radiation contribution to the solution defined in Eq.~\eqref{eq:radiation} and can be obtained from  Eq.~\eqref{eq:QCsoln_subkinks} and Eq.~\eqref{eq:QCsoln_shocks} for subkinks and shocks, respectively. The difference is that for subkinks  the function $G(V)$ is known once and for all, while for shocks we can only obtain a one-parametric family of such functions.

\subsection{Admissibility}
\begin{figure}[h!]
  \centering
      \center{\includegraphics[scale=0.4]{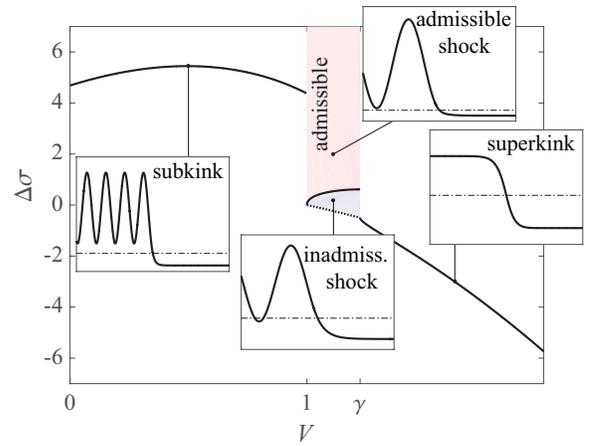}}
  \caption{Admissibility sets of solutions of the QC problem. In the blue region we have $\varepsilon(\eta)\leq \varepsilon_c$ for some intervals of $\eta<\varepsilon_c$, and its dotted lower boundary marks the threshold $\varepsilon_{-}=\varepsilon_c$. The insets show examples of the strains $\varepsilon(\eta)$, with horizontal lines marking $\varepsilon=\varepsilon_c$. Here $\gamma^2=1.5$, $\varepsilon_c=1$, and we set $\varepsilon_{+}=0$.}
\label{fig:admissibilityQC}
\end{figure}
We recall that the explicit expressions for the general solution of the piecewise linear problem are invalid if the admissibility conditions \eqref{eq:admissibility} are violated. Therefore the inequalities
 $\varepsilon(\eta)>\varepsilon_c$ for $ \eta<0$ and
 $\varepsilon(\eta)<\varepsilon_c$ for $\eta>0$ must be checked a posteriori, which means that some of the formally constructed solutions may have to be  discarded \cite{atkinson1965motion,marder1995origin}.

The analysis of the  global behavior of the obtained  strain fields shows that all subkinks with $V<1$ and all superkinks with $V>\gamma$ are automatically admissible. In both of these these cases the transition fronts can be represented in the space of parameters $\varepsilon_+$ and $\Delta \sigma$ by one-dimensional manifolds  because the velocity of the front is determined uniquely by the corresponding kinetic relation. In the case of shocks, which can be either admissible or inadmissible, the velocity $V$ is not determined internally. Therefore, shocks occupy a two-dimensional (2D) domain in the $(\e_+,\,\Delta \sigma)$ plane. This domain is further divided into two subdomains: at sufficiently large values of $\Delta \sigma$ shocks are admissible, while those  located below a certain threshold are inadmissible. The inadmissible shocks show  the repeated crossing of the $\varepsilon_c$ threshold by the oscillatory tail  behind the moving front.

The admissibility diagram in $(V,\,\Delta \sigma)$ plane is shown in Fig.~\ref{fig:admissibilityQC}, where we fixed $\varepsilon_+=0$.  The insets illustrate  the  analytical solutions describing different types of transition fronts. The 2D domain of shocks on this diagram is bounded on two sides by the condition $1<V<\gamma$ and from below by the dotted line below which $\e_{-}<\e_c$. One can see that
only the shock solutions in the pink (upper) region above the threshold values $\Delta\sigma^*(V)$ marked by a solid black line are admissible, while the ones in the blue (lower) region are inadmissible. This is illustrated on the corresponding inset by the multiple crossings of $\e_c$ (the dash-dotted horizontal line) by the strain profile $\e(\eta)$.

To understand which solutions replace shocks in the `forbidden' region, we need to resort to simulations. Using direct numerical simulations of Eq.~\eqref{eq:QCeqn_displ} and using a sufficiently broad set of initial data we can also numerically test the stability of the admissible transition fronts.

\begin{figure}[h!]
  \centering
  \begin{tabular}{@{}p{0.45 \linewidth}@{\quad}p{0.45 \linewidth}@{}}
    \subfigimg[width=\linewidth]{(a)}{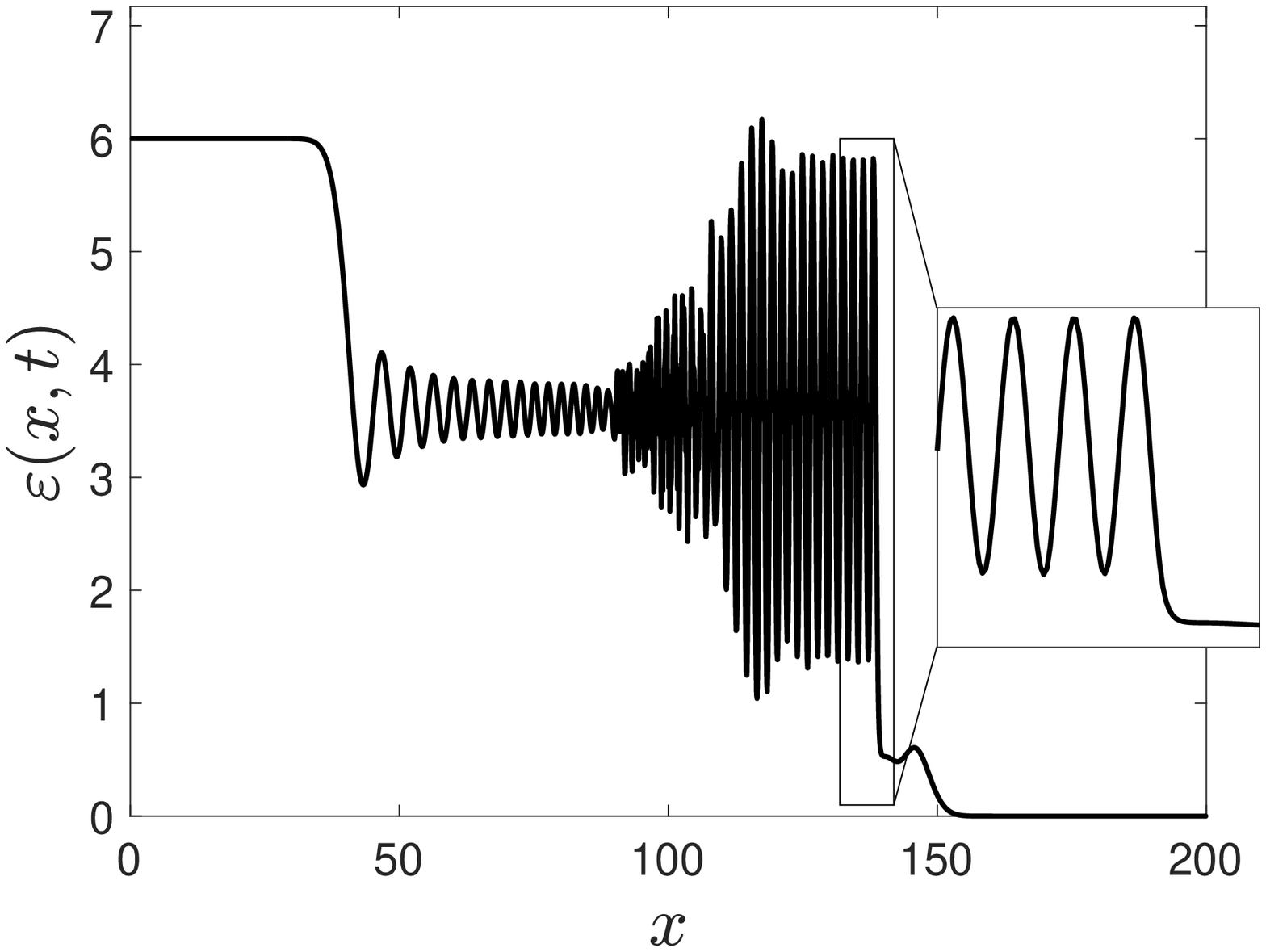} &
    \subfigimg[width=\linewidth]{(b)}{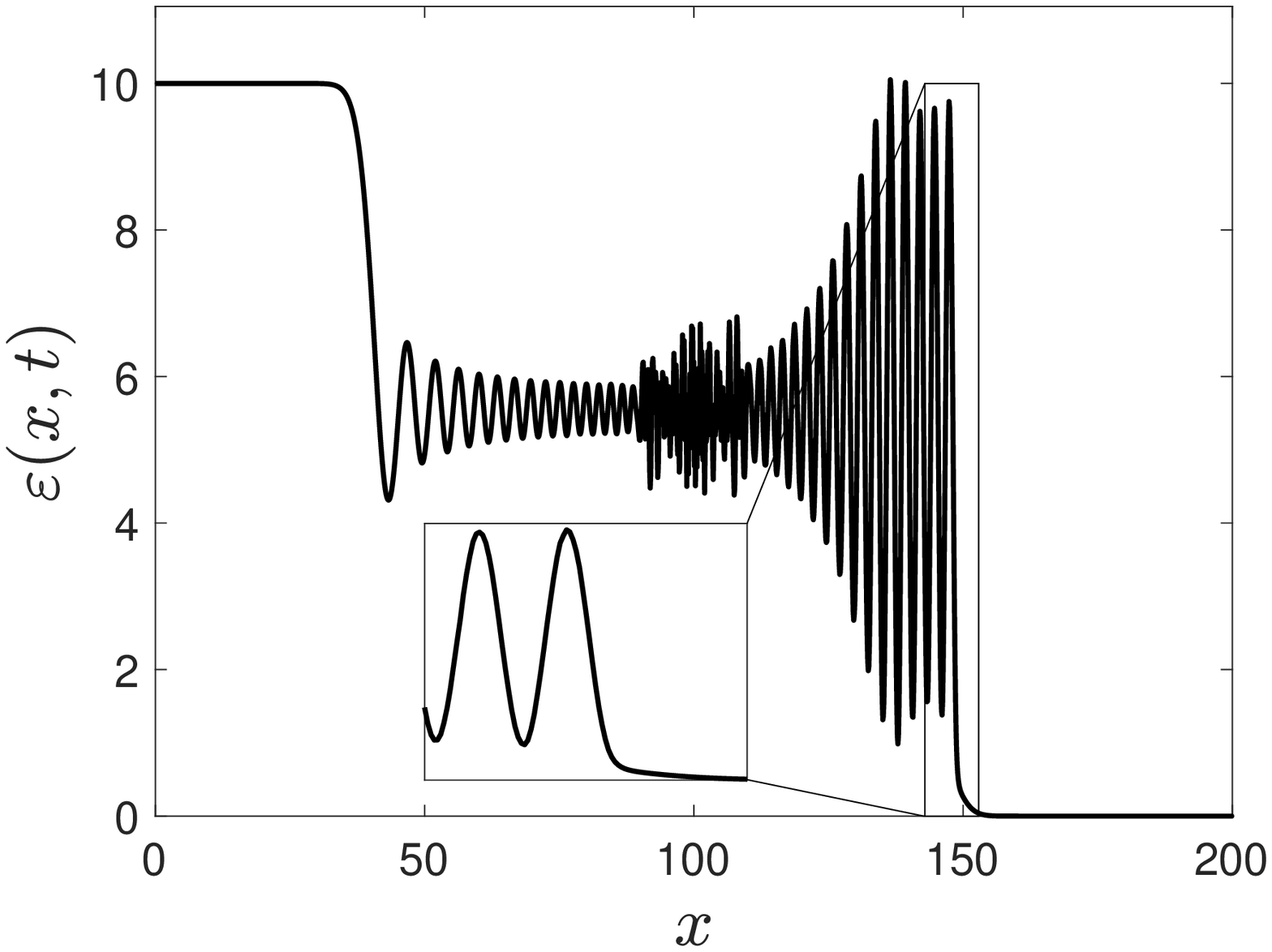} \\
    \subfigimg[width=\linewidth]{(c)}{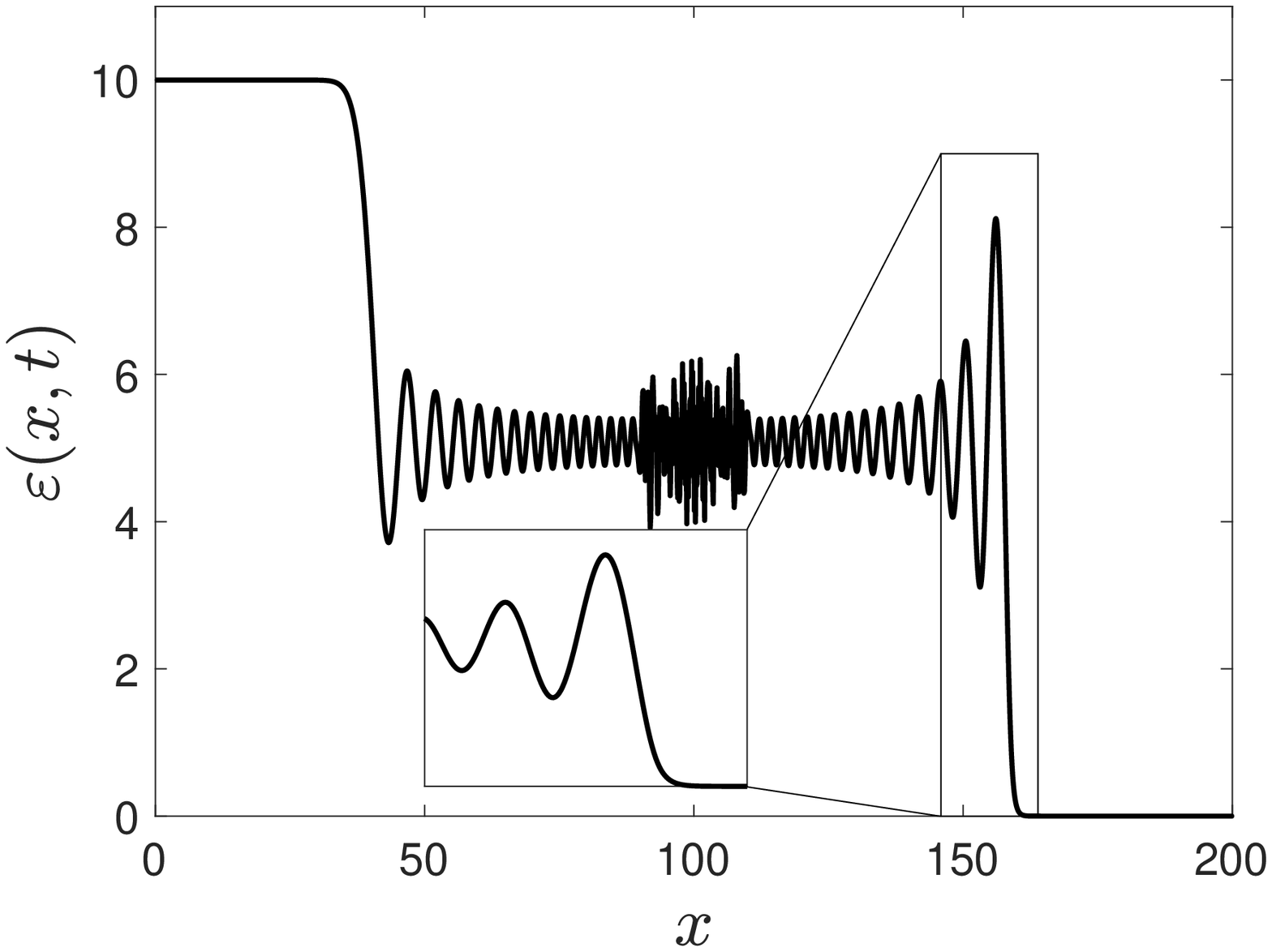} &
    \subfigimg[width=\linewidth]{(d)}{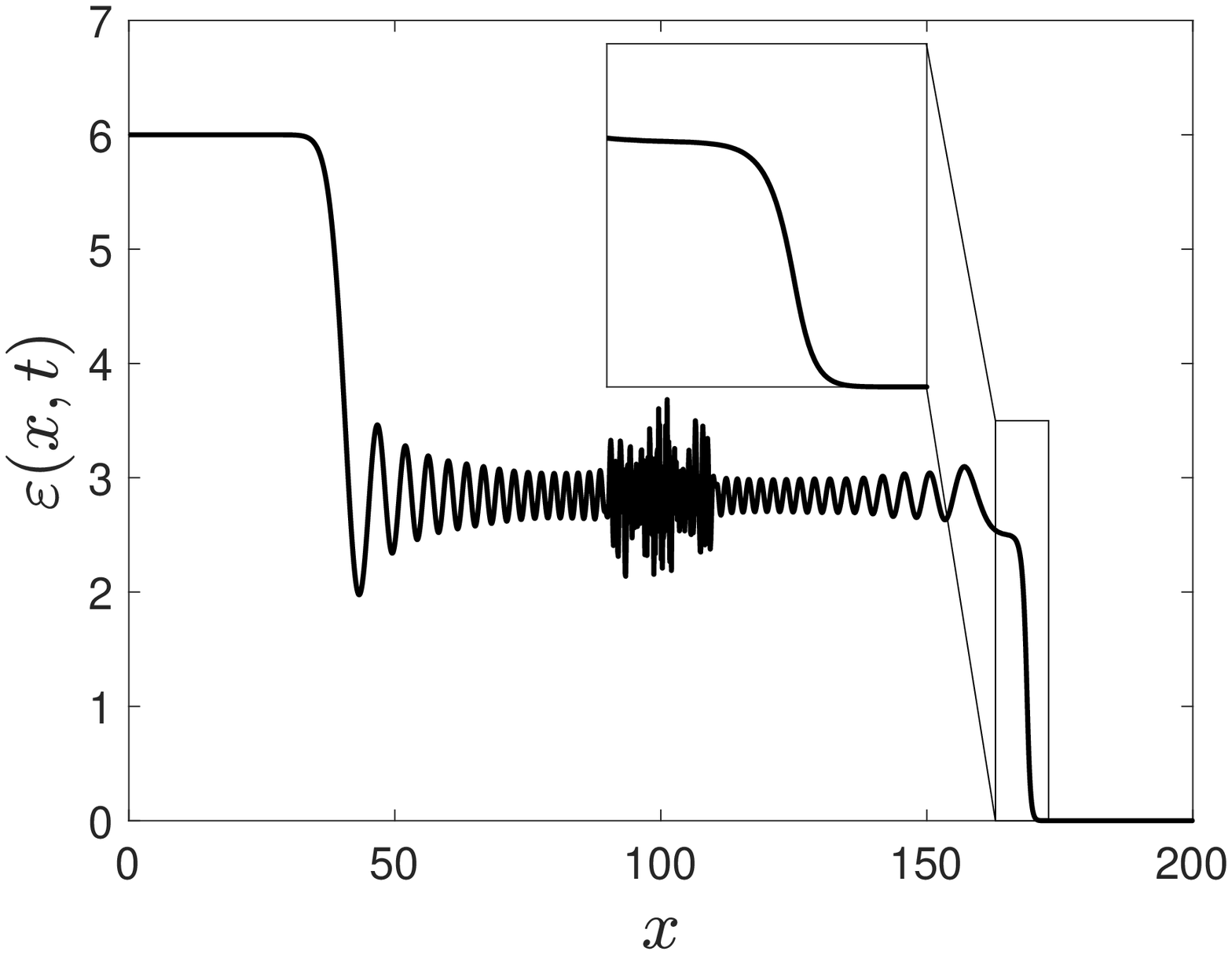}
  \end{tabular}
  \caption{Different regimes of front propagation in the QC model with the parameters $\gamma^2=1.5$ and $\varepsilon_c=1$ at $t=50$: (a) subkink ($\varepsilon_l=6$, $\Delta\sigma=2.5$); (b) conventional shock ($\varepsilon_l=10$, $\Delta\sigma=2.5$); (c) dispersive shock ($\varepsilon_l=10$, $\Delta\sigma=0$); (d) superkink ($\varepsilon_l=6$, $\Delta\sigma=-1.5$).}
\label{fig:numerics_quasi}
\end{figure}
\subsection{Numerical simulations}
We solve Eq.~\eqref{eq:QCeqn_strain} in the finite domain $x\in(0,200)$  with the Riemann-type initial data
\[
\varepsilon(x,0)=\begin{cases}
\varepsilon_l,\,x<100,\\
0,\,x\geq 100,
\end{cases}\quad \frac{\partial \varepsilon}{\partial t}(x,0)=0.
\]
using the implicit fourth-order conservative finite-difference method developed in \cite{Wang2018}. The first and second spatial derivatives of strain are set to zero at the boundaries. The emergence of particular transition fronts, as an outcome of the breakdown of the unstable initial state, will then depend on the choice of the parameters $\Delta\sigma$ and $\varepsilon_l$.

The results are summarized in Fig.~\ref{fig:numerics_quasi}, which shows time snapshots near the end of four different simulations.  In each simulation we have chosen a particular set of  parameters $\varepsilon_l$ and $\Delta\sigma$ to reach one of the four structurally dissimilar regimes shown in  Fig.~\ref{fig:admissibilityQC}.

While in all presented snapshots we observe complex breakdown patterns, most of their elements correspond to linear dispersive pulses with their characteristic overshoots. To identify genuinely nonlinear substructures one needs to look for the patterns magnified in the insets in Fig.~\ref{fig:admissibilityQC}. Thus, the inset in Fig.~\ref{fig:numerics_quasi}(a) shows an admissible subkink moving to the right. The comparison of the internal structure of such numerically generated  wave profile  with the corresponding analytical solution shows perfect agreement, which  confirms that  the transformation fronts of this type can indeed serve as dynamical attractors.   Similarly, the inset in Fig.~\ref{fig:numerics_quasi}(d) shows an admissible  superkink moving to the right,  which  also matches the analytical waveform and points towards  stability of the corresponding traveling wave solution. An admissible shock is shown in the inset of Fig.~\ref{fig:numerics_quasi}(b), and we again see that the analytical profile is reproduced faithfully and conclude that such transition fronts can be stable. The remaining panel (c) of Fig.~\ref{fig:numerics_quasi} corresponds to parameter values that target inadmissible shocks. Not surprisingly, we do not observe a traveling wave profile in this case. Instead, the nonlinear structure that we see is reminiscent of a non-steady dispersive shock wave (DSW).

Our broader numerical experiments strongly suggest that, in the whole domain of non-admissibility, shock traveling waves are replaced by DSWs. This result, obtained so far only in the QC setting, will be confirmed below by a similar analysis of the original discrete problem.  We recall that DSWs have been extensively studied using various other QC approximations of the FPU system (see, for example, \cite{gurevich1973nonstationary,congy2019nonlinear,ablowitz2013dispersive,kamchatnov2019dispersive,benzoni2021modulated}).
We can then conclude that in our regime diagram shown in Fig.~\ref{fig:admissibilityQC} the domain of inadmissible shocks  should be interpreted as a domain of stability of DSW type non-steady (spreading) transition fronts. The absence of steadily moving shock fronts in the FPU model with convex energy density  ($\Delta \sigma\leq0$ in our problem) is well known. It has been previously linked to the low dimensionality (lack of transversal radiation) and the absence of irreversibility (purely elastic constitutive modeling), which is a ubiquitous feature of the real crystals \cite{holian1995atomistic,zhakhovskiui1999shock,stoltz2005shock}. Here by allowing regimes with  $\Delta \sigma\geq0$ we acquire a limited parametric domain where stable stationary shocks exist.  One can argue that the implied nonconvexity, which allows the system to accommodate large-amplitude lattice waves  transmitting radiated energy away from the moving front, is the way to bring multivaluedness into the  the constitutive response, which ultimately imitates the inherent multistability of the plastic response. That one-dimensional shock traveling waves are not possible under the assumption of energy convexity is confirmed in our numerical simulation results by the appearance of the inadmissible region in the regime diagram, where the steadily moving transition fronts  are replaced by the spreading DSW profiles.

To summarize, the analysis of the dispersively regularized QC model allowed us to clarify the ambiguities left by the classical continuum description.  In such essentially microscopic  model all three classes of transition fronts acquired their natural raison d'\^etre, with the numerical simulations providing confirmation of the exhaustiveness for the proposed classification. It is rather remarkable that such a task could be accomplished using a relatively simple QC approximation of the original discrete problem.  Note, however, that the chosen approximation was not of the lowest order, and to capture the complete picture we had to introduce two internal time scales and modify the kinetic rather than elastic energy. As we show in the next section, the obtained  description is fully adequate when compared to the results discussed below for the discrete model.

\section{Discrete model}
\label{Discrete model}
We now analyze the dimensionless version of the original FPU problem \eqref{eq:EoM_strain}, which takes the form
\beq
\frac{d^2 \e_n(t)}{dt^2}=\sigma(\e_{n+1})-2\sigma(\e_n)+\sigma(\e_{n-1}),
\label{eq:discrete_dyn}
\eeq
with bilinear interactions $\sigma(\e)=\e$ at $\e<\e_c$, and $\sigma(\e)=\gamma^2\e-\sigma_0$ at $\e>\e_c$. The dispersion relations in each linear regime are defined by
\beq
\omega_+^2(k)=\omega_-^2(k)/\gamma^2= 4\sin^2\left( k/2\right)
\label{eq:dispersion_D}
\eeq
and are much more intricate than in the QC model due to the presence of lattice resonances and the richness of the spectrum of available lattice-scale waves. Therefore the analysis of the discrete problem can potentially challenge the description of the energy radiation provided by the QC  model.

To find the corresponding traveling waves solutions $\e_n(t)=\e(\eta)$, $\eta=n-Vt$, of the discrete problem \eqref{eq:discrete_dyn}, we need to solve the advance-delay equation
\begin{equation}
V^2\frac{d^2\varepsilon}{d\eta^2}=\sigma(\eta+1)+\sigma(\eta-1)-2\sigma(\eta),
\label{eq:equations_steady}
\end{equation}
where the function $\sigma(\eta)=\sigma(\e(\eta))$ is given by Eq.~\eqref{QCeqn2}. We will use Fourier transform technique to solve Eq.~\eqref{eq:equations_steady} subject to the consistency condition \eqref{eq:consistency}, the boundary conditions \eqref{eq:BCs} and the radiation conditions \eqref{eq:rad_conds}.

\begin{figure}[h!]
  \centering
  \begin{tabular}{@{}p{0.48 \linewidth}@{\quad}p{0.48 \linewidth}@{}}
    \subfigimg[width=\linewidth]{(a)}{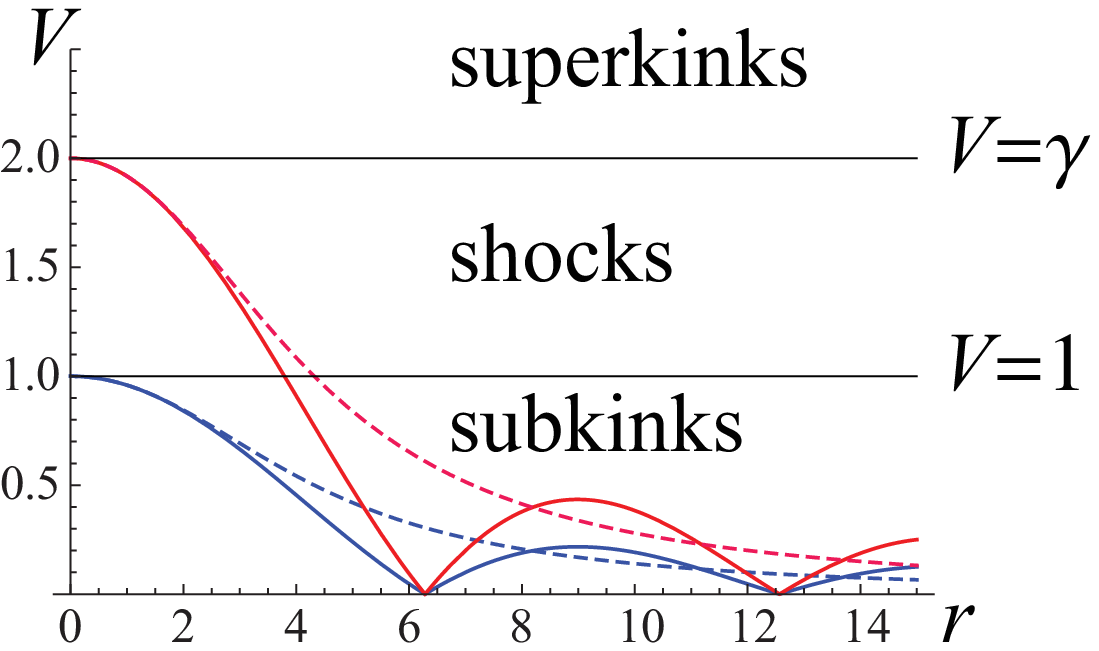} &
    \subfigimg[width=\linewidth]{(b)}{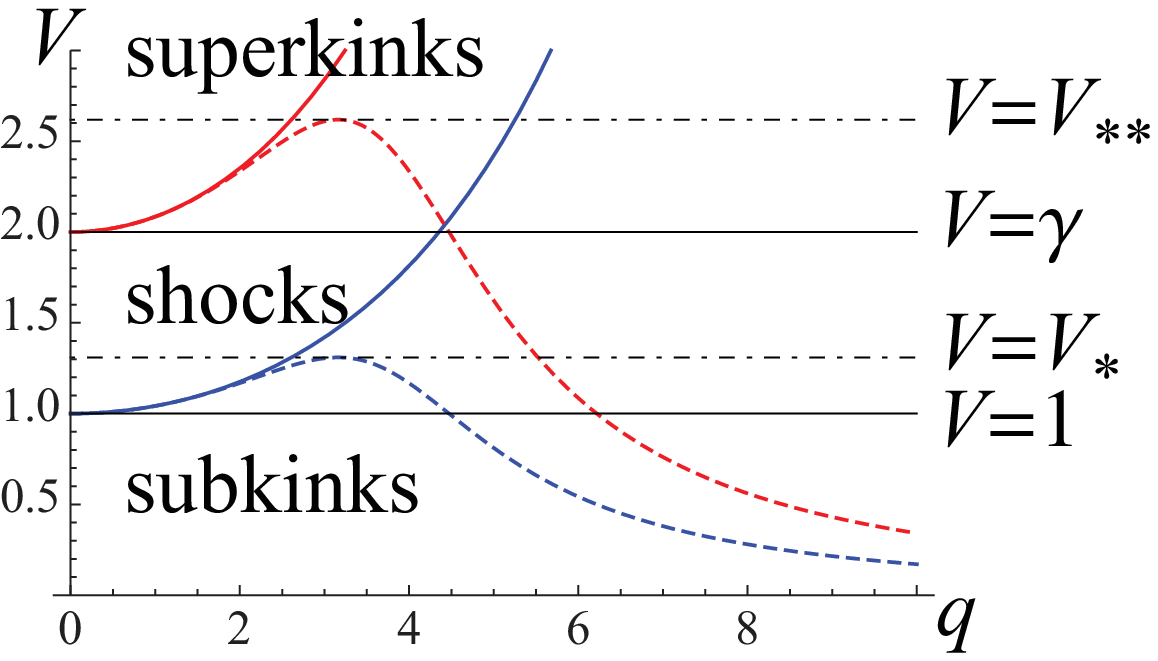}
  \end{tabular}
  \caption{The structure of the magnitudes of (a) the real roots $k=\pm r$ and (b) the imaginary roots $k=\pm iq$ in the discrete (solid curves) and QC (dashed curves) for $\omega^2_{+}(k)-k^2V^2=0$ (blue curves) and $\omega^2_{-}(k)-k^2V^2=0$ (red curves). Black dashed curves mark the sonic limits separating the velocity domains of different transition fronts. Complex roots with nonzero real and imaginary parts bifurcate at the velocities $V_*$ and $V_{**}$ marked by dash-dotted lines in the QC model and from the non-sonic maxima of the real root curve in (a) in the discrete model. Here $\gamma=2$.}
\label{fig:roots_DvsQC}
\end{figure}

It is convenient to represent the transformed function in the form
\[
\begin{gathered}
\hat{\varepsilon}(k)=\int_{-\infty}^{\infty}\varepsilon(\eta)e^{ik\eta}\,d\eta=\hat{\varepsilon}^+(k)
+\hat{\varepsilon}^-(k),
\end{gathered}
\]
where
\[
\hat{\varepsilon}^\pm(k)=\int_{-\infty}^{\infty}\varepsilon(\eta)H(\pm\eta)e^{ik\eta}\,d\eta
\]
are analytic in $\Im k \gtrless 0$.
The  Fourier transform of \eqref{eq:equations_steady} then yields
\begin{equation}
M_+\hat{\varepsilon}^{+}+ M_-\hat{\varepsilon}^{-}=
\frac{M_{-}-M_{+}}{ik}\varepsilon_*,
\label{eq:WH}
\end{equation}
where we introduced the parameter
\beq
\varepsilon_*= \dfrac{\sigma_0}{\gamma^2-1}
\label{eq:eps_star}
\eeq
and the characteristic functions
 \beq
M_{\pm}(k)=\omega_{\pm}^2(k)+(0+ikV)^2.
\label{eq:Lpm_def}
\eeq
Here $0\pm ikV=\lim_{s\to 0+}(s \pm ikV)$, and we use the causality principle \cite{slepyan2012models} to handle the zero at the origin. A comparison of the characteristic functions \eqref{eq:Lpm_def} with their QC analogs in the whole complex plane shows that while the discrete dispersion relations \eqref{eq:dispersion_D} are more complex than their QC counterparts  \eqref{eq:dispersion_QC}, the QC approximation captures the long-wave behavior adequately.  More precisely, as shown in Fig.~\ref{fig:roots_DvsQC}, the QC model gives an excellent approximation of the real and purely imaginary roots of Eq.~\eqref{eq:Lpm_def} that have sufficiently small magnitude. In general, it captures the four nonzero roots of each characteristic function that are closest to $k=0$ qualitatively well but may represent purely imaginary roots by complex quadruples and vice versa.

\subsection{Characteristic roots}
\label{sec:roots}
\begin{figure*}
  \centering
  \begin{tabular}{@{}p{0.3\linewidth}@{\quad}p{0.3\linewidth}@{\quad}p{0.3\linewidth}@{}}
    \subfigimg[width=\linewidth]{(a)}{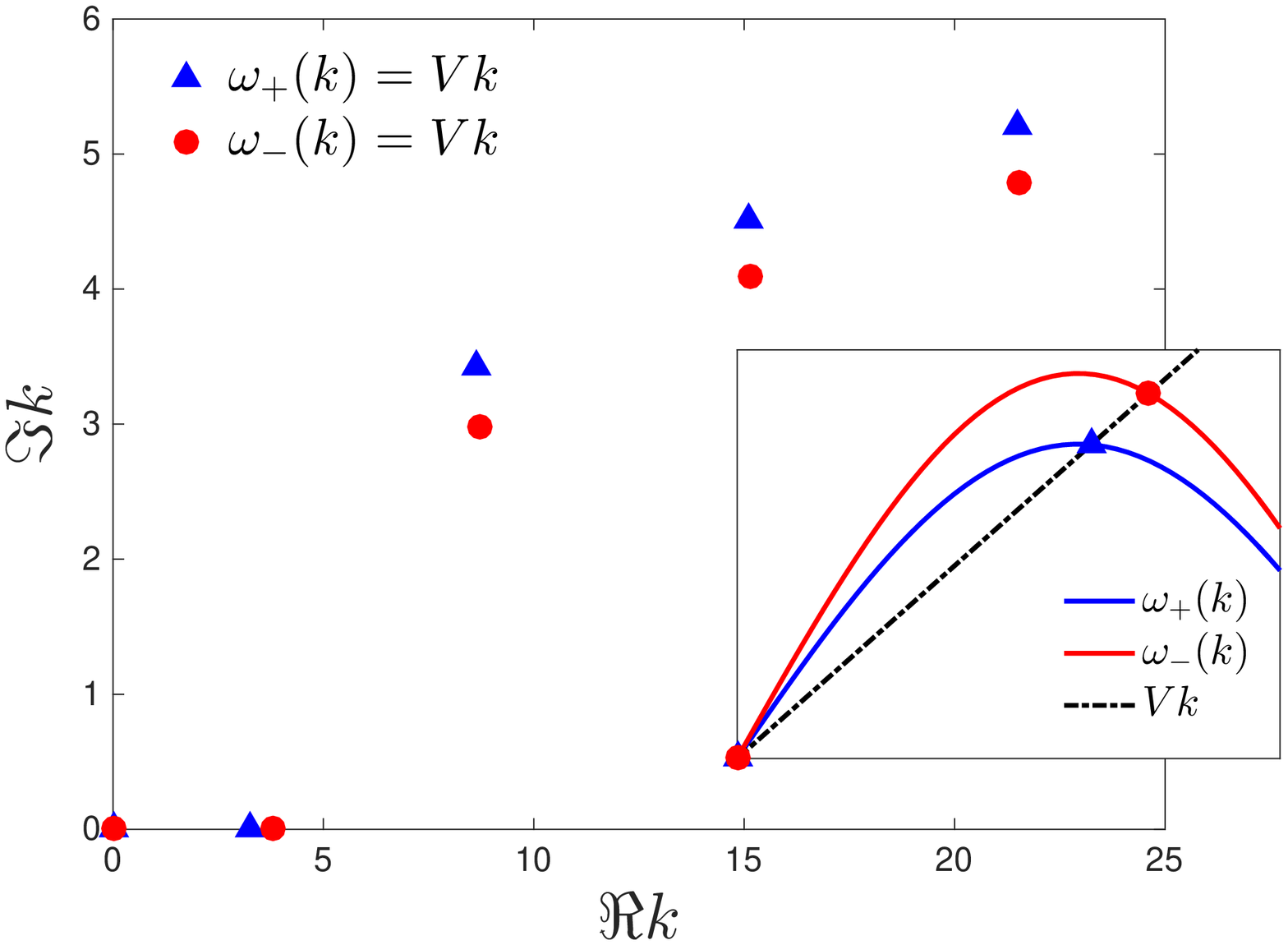} &
    \subfigimg[width=\linewidth]{(b)}{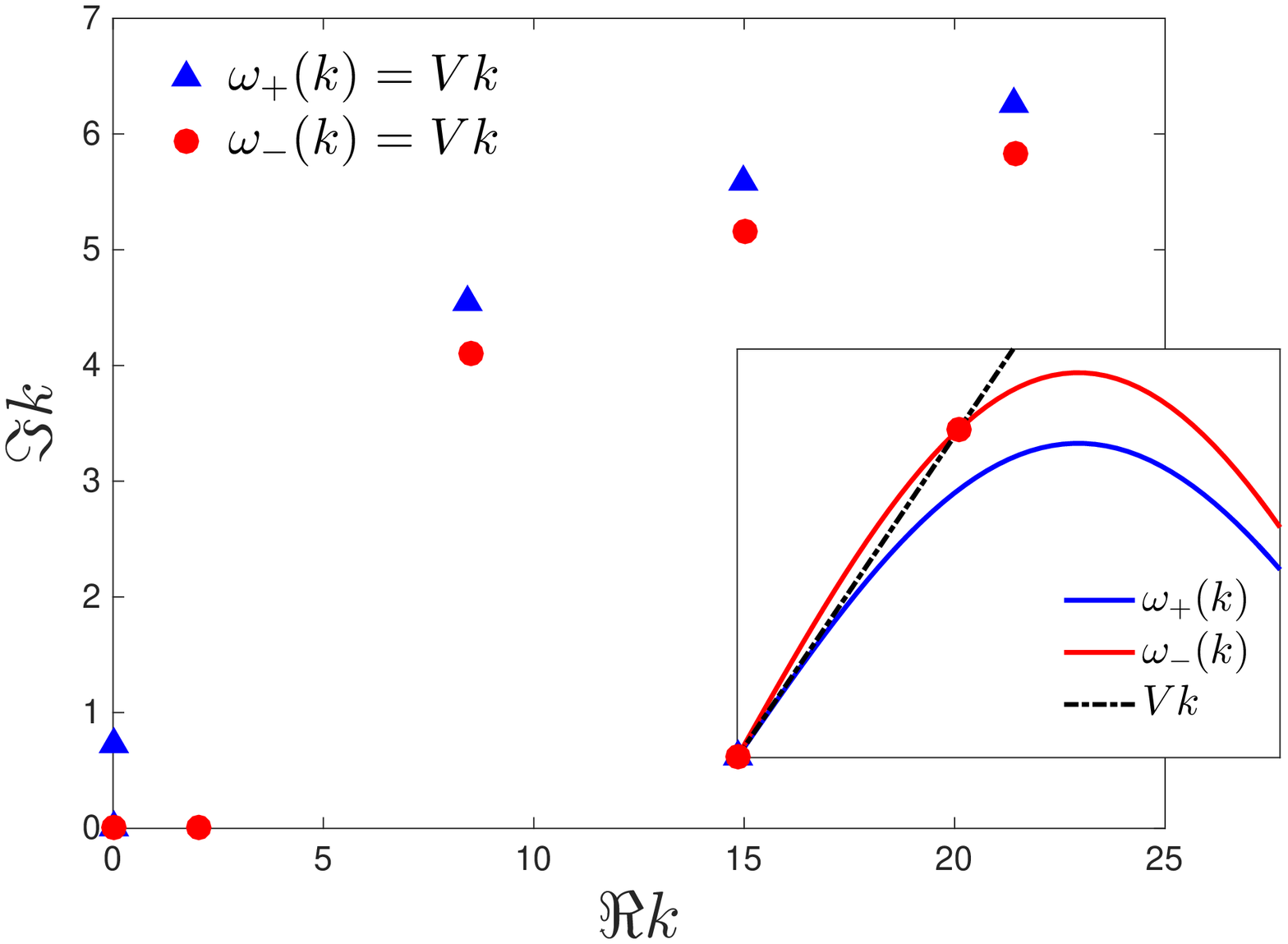} &
    \subfigimg[width=\linewidth]{(c)}{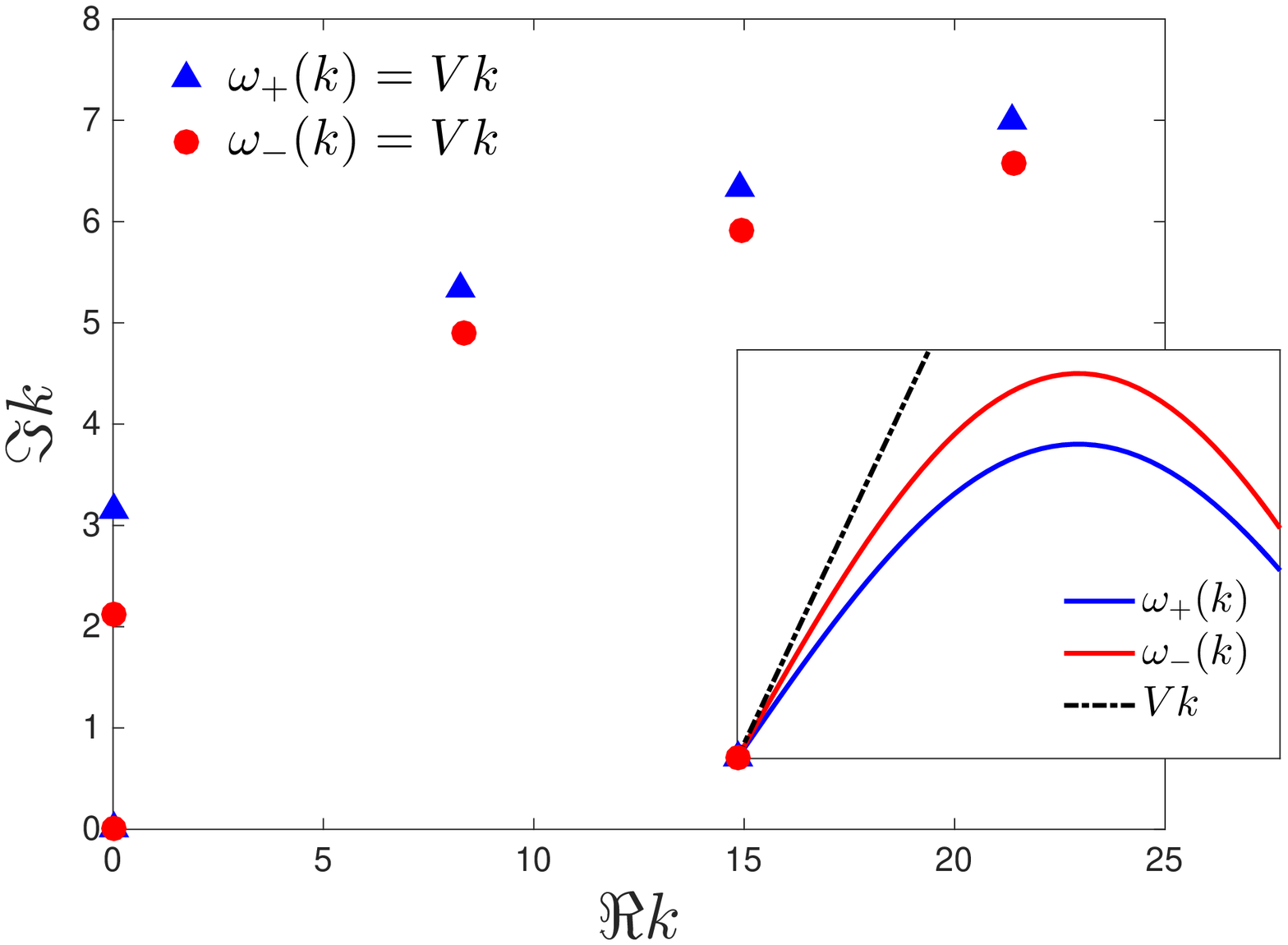}
  \end{tabular}
  \caption{Distribution of the roots of $M_{\pm}(k)$ in Eq.~\eqref{eq:Lpm_def} for the discrete model when (a) $V<1$, (b) $1<V<\gamma$, (c) $V>\gamma$. Due to symmetry, only the roots with $\Re k \geq 0$ and $\Im k \geq 0$ are shown. Insets show the dispersion relations and real roots as intersections with the line $Vk$.}
\label{fig:Roots}
\end{figure*}
Similar to the QC model, the solution of the discrete problem can be written in terms of elementary waveforms associated with the roots of the characteristic
functions \eqref{eq:Lpm_def}. In what follows, we consider the generic case when $V$ is non-resonant ($V \neq \omega_+'(k)$ and $V \neq \omega_-'(k)$ for any real $k$). We can then define the sets ${\cal Z} = {\cal Z}_r^+\cup{\cal Z}_r^-\cup {\cal Z}_c^+\cup{\cal Z}_c^-$ and ${\cal P} = {\cal P}_r^+\cup{\cal P}_r^-\cup {\cal P}_c^+\cup{\cal P}_c^-$ containing  nonzero roots of the characteristic equations $M_{\pm}(k)=0$. Here
\begin{equation}
\begin{split}
&{\cal Z}_r^\pm=\{z : M_+(z)=0,\, z \neq 0, \, \Im z=0,\,\omega_+'(z) \gtrless V\},\\
&{\cal P}_r^\pm=\{p: M_-(p)=0,\, p \neq 0,\, \Im p=0,\,\omega_-'(p)
\gtrless V\},\\
&{\cal Z}_c^\pm=\{z : M_+(z)=0,\, \Im z \lessgtr 0\},\\
&{\cal P}_c^\pm=\{p: M_-(p)=0,\,  \Im p \lessgtr 0\}.
\end{split}
\label{eq:root_sets}
\end{equation}
The structure of the roots of Eq.~\eqref{eq:Lpm_def} is illustrated in Fig.~\ref{fig:Roots}, which can be compared to the corresponding root structure for the QC model shown in Fig.~\ref{fig:Roots_quasi} (see also Fig.~\ref{fig:roots_DvsQC}, which compares the structure of real and purely imaginary roots). As in that case, the even symmetry of each characteristic function implies that the roots are symmetric about the origin, and it suffices to consider the region $\Re k \geq 0$ and $\Im k \geq 0$.

Of particular importance are the sets of nonzero real roots ${\cal Z}_r^+\cup{\cal Z}_r^{-}$ (roots of $M_+(k)$) and
${\cal P}_r^+\cup{\cal P}_r^{-}$ (roots of $M_{-}(k)$). As we will see, some of these roots correspond to radiated lattice waves. When the sets are nonempty for given non-resonant $V$, they contain an odd number of positive real roots, given by $2l+1$ and $2m+1$, respectively. We arrange these roots in the ascending order: $z_j<z_{j+1}$, $j=1,\dots,2l$, and $p_j<p_{j+1}$, $j=1,\dots,2m$.

We observe that in the case of superkinks ($V>\gamma$) both functions $M_{\pm}(k)$ have no nonzero real roots, as shown in Fig.~\ref{fig:roots_DvsQC}(a) and Fig.~\ref{fig:Roots}(c), and hence there are no radiated waves in this case (no dissipation). For shocks ($1<V<\gamma$) only $M_-(k)$ has such roots (see Fig.~\ref{fig:roots_DvsQC}(a) and Fig.~\ref{fig:Roots}(b)). More specifically, we have $m=0$ (i.e., one positive real root) for the values of velocity $V$ below the first resonance velocity $V_1$ which solves $\omega_{-}'(k)=V_1 k$ for some real $k$ and naturally satisfies the condition $V_1>1$.  We then have  $m=1$ (three positive real roots) for the values of $V$ between the first and second resonance velocities, where the second resonance velocity is defined accordingly, and so on. Finally, for subkinks ($V<1$) each of the characteristic equations has at least one positive real root (see Fig.~\ref{fig:roots_DvsQC}(a) and Fig.~\ref{fig:Roots}(a)), with $l$ and $m$ each increasing by one when the corresponding resonance velocity is crossed.

In addition to real roots, there are infinite sets of complex roots ${\cal Z}_c^+\cup{\cal Z}_c^{-}$ (roots of $M_+(k)$) and
${\cal P}_c^+\cup{\cal P}_c^{-}$ (roots of $M_{-}(k)$) with nonzero imaginary part that can be seen in Fig.~\ref{fig:Roots}. These roots bifurcate from the maxima of the real-root curves shown in Fig.~\ref{fig:roots_DvsQC}(a). This includes purely imaginary roots that bifurcate from the sonic maxima at $k=0$ and are shown in Fig.~\ref{fig:roots_DvsQC}(b). The non-real roots define the structure of the boundary layers on both sides of the moving front.

\subsection{Characteristics revisited}
To make a connection with the classical continuum theory, we recall that the configuration of the real roots $z_j$ and $p_j$ around the origin $k=0$ is intimately related to the structure of the characteristics in the continuum approximation.  Therefore by studying these roots one can expect  to reconstruct  the main subdivision of the transformation fronts into the three universality classes.

We shall exploit the fact that in the long-wavelength limit the discrete problem can be replaced by a single nonlinear wave equation. Indeed, in the limit
$k \to 0$, $s\to 0+$  we  can  approximate the linear operators in Eq.~\eqref{eq:Lpm_def} by
\begin{equation}
\begin{split}
&M_+(k)=\omega_+^2(k)+(s+ikV)^2\approx g_+(k,s)\\
&\equiv \left((1+V)(-ik)-s\right)\left((1-V)(-ik)+s\right),\\
&M_-(k)=\omega_-^2(k)+(s+ikV)^2\approx g_-(k,s)\\
&\equiv \left((\gamma+V)(-ik)-s\right)\left((\gamma-V)(-ik)+s\right),
\end{split}
\label{eq:Dispersions_zero}
\end{equation}
Observe also that using the convective coordinate $\eta=x-Vt$  we can  rewrite the system \eqref{eq:cont} as a pair of linear wave equations for $\varepsilon(\eta,t)$ in each of the two domains of linearity:
\begin{equation}
\begin{split}
&\left[(1+V)\frac{\partial\e}{\partial \eta}-\frac{\partial\e}{\partial t}\right]\left[(1-V)\frac{\partial\e}{\partial \eta}+\frac{\partial\e}{\partial t}\right]=0,\quad \eta>0,\\
&\left[(\gamma+V)\frac{\partial\e}{\partial \eta}-\frac{\partial\e}{\partial t}\right]\left[(\gamma-V)\frac{\partial\e}{\partial \eta}+\frac{\partial\e}{\partial t}\right]=0,\quad \eta<0.
\end{split}
\label{eq:Product_Macro}
\end{equation}
Applying Fourier transform in $\eta$  and Laplace transform in $t$  transforms Eq.~\eqref{eq:Product_Macro} into the equations $g_{\pm}(k,s)=0$, where the functions $g_{\pm}(k,s)$ are defined in Eq.~\eqref{eq:Dispersions_zero}.

Since the characteristics of Eq.~\eqref{eq:Product_Macro} are defined by the equations
$\eta\pm(1\pm V)t=\text{const}$ at $\eta>0$ and $\eta\pm(\gamma \pm V)t=\text{const}$ at $\eta<0$, the location of the roots of the functions $g_{\pm}(k,0)$
is directly linked to the configuration of the characteristics relative to the line $\eta=\text{const}$.
 \begin{figure}[h!]
  \centering
  \begin{tabular}{@{}p{0.15 \linewidth}@{\quad}p{0.15 \linewidth}@{\quad}p{0.15 \linewidth}@{}}
    \subfigimg[width=\linewidth]{(a)}{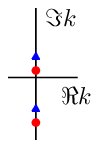} &
    \subfigimg[width=\linewidth]{(b)}{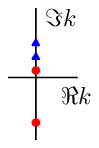} &
    \subfigimg[width=\linewidth]{(c)}{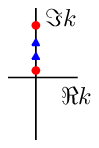}
  \end{tabular}
  \caption{Schematic presentation of the roots $g_+(k,0)$ (blue triangles) and $g_-(k,0)$ (red circles): (a) $V<1$, (b) $1<V<\gamma$, (c) $V>\gamma$.}
\label{fig:Roots_characteristics}
\end{figure}
The configuration of the roots of the equations $g_{\pm}(k,0)=0$ is shown schematically in Fig.~\ref{fig:Roots_characteristics} separately for each class  of the transition fronts.  One can see that in the range $V<1$ (subkinks) the purely imaginary roots are located in two different complex half-planes for both $g_{+}(k,0)=0$ and $g_{-}(k,0)=0$. This is equivalent to the fact that there is one incoming and one outgoing characteristic on both sides of  the line $x-Vt=\text{const}$. Both roots of the equation $g_+(k,0)=0$ end up in the upper complex half-plane in the range  $1<V<\gamma$ (shocks), producing two incoming characteristics on the right side of the line $x-Vt=\text{const}$, while there is still one incoming and one outgoing characteristic on the left side.  Finally, in the range $V>\gamma$ (superkinks), the remaining roots of $g_-(k,0)=0$ also shift into the upper complex half-plane, which produces  two outgoing characteristics behind the moving front. One can see that the location of the roots in Fig.~\ref{fig:Roots_characteristics} is in full agreement with the propagation direction of the macroscopic perturbations with respect to the moving front for each of our universality classes, as shown in Fig.~\ref{fig:Characteristics}.

\subsection{Solution of the discrete problem}
We observe that $\hat{\e}^{\pm}(k)$ can be written as
\begin{equation}
\hat{\e}^\pm(k)=\dfrac{\varepsilon_{\pm}}{0\mp ik}+\hat{\chi}^\pm(k),
\label{eq:chi0}
\end{equation}
where the first term accounts for the boundary conditions \eqref{eq:BCs}, and the second term satisfies $\lim_{k \to \pm i0}\hat{\chi}^{\pm}(k)=0$, so that $\lim_{\eta \to \pm \infty} \langle \e(\eta) \rangle=\lim_{k \to \pm i0}\hat{\e}^{\pm}(k)=\e_{\pm}$.

To find $\hat{\chi}^\pm(k)$, we use the Wiener-Hopf technique \cite{slepyan1984fracture,slepyan2005transition,trofimov2010shocks,kresse2004lattice}.
To this end, we factorize the main linear operator
\beq
L(k)=\dfrac{M_+(k)}{M_-(k)}=\dfrac{\omega_+^2(k)+(0+ikV)^2}{\omega_-^2(k)+(0+ikV)^2}
\label{eq:L_def}
\eeq
of  the problem, which means representing it in the form
\beq
L(k)=L^+(k)L^-(k),
\label{eq:Lfactor}
\eeq
where the superscripts $\pm$ identify functions that are regular (have no zeroes or singularities) in $\Im k \gtrless 0$, respectively. Such factorization allows us to rewrite \eqref{eq:WH} as
\begin{equation}
\begin{split}
&L^+(k)\left[-\varepsilon_*-ik\left(\hat{\chi}^+(k)+\dfrac{\varepsilon_{+}}{0-ik}\right)\right]\\
&=\frac{1}{L^-(k)}\left[ik\left(\hat{\chi}^-(k)+\dfrac{\varepsilon_{-}}{0+ik}\right)-\varepsilon_*\right].
\end{split}
\label{eq:WH_second}
\end{equation}
This representation ensures that the right hand side is regular in the lower half-plane, while the left hand-side is regular in the upper half-plane, so that both can be analytically continued to the whole plane after we move the zeroes and singularities along the real axis into the corresponding half-planes.

Using the infinite product theorem \cite{noble1958methods} we can represent $L^{\pm}(k)$ as follows \cite{slepyan1982antiplane}:
\beq
L^{\pm}(k)=l^{\pm}(k) {L_0}^{\pm}(k).
\label{eq:Lfactor_more}
\eeq
Here the terms $l^{\pm}(k)$ depend on nonzero real roots of the characteristic equations, while the terms ${L_0}^{\pm}(k)$ are defined by the remaining non-real (complex) roots.

More specifically, we have
\beq
L_0^{\pm}(k)=\sqrt{\dfrac{1-V^2}{\gamma^2-V^2}}\dfrac{\prod\limits_{z \in {\cal Z}_c^\pm}\left(1-\dfrac{k}{z}\right)}
{\prod\limits_{p \in {\cal P}_c^\pm}\left(1-\dfrac{k}{p}\right)}
\label{eq:L0pm}
\eeq
where the products are over the sets ${\cal Z}_c^{\pm}$ and ${\cal P}_c^{\pm}$ of non-real roots defined in Eq.~\eqref{eq:root_sets}. Note that the zeroes and poles of $L_0^+(k)$ (the set ${\cal Z}_c^+ \cup {\cal P}_c^+$) are all located in $\Im k<0$, and the zeroes and poles of of $L_0^-(k)$ (the set ${\cal Z}_c^- \cup {\cal P}_c^-$) are all in $\Im k>0$.

Similarly, the functions $l^{\pm}(k)$ can be expressed in terms of the nonzero real roots of the corresponding characteristic equations belonging to the sets ${\cal Z}_r^{\pm}$ and ${\cal P}_r^{\pm}$ in Eq.~\eqref{eq:root_sets}. These roots are placed into the ``+'' sets (which contribute to the solution at $\eta>0$) if the associated group velocities $\omega'(k)$ exceed the phase velocity $V$ and into ``-'' sets (contributing to the solution at $\eta<0$) if $\omega'(k)<V$. This ensures that the solution satisfies the radiation condition \eqref{eq:rad_conds}, and the radiated waves carry energy away from the front. Recalling the the structure of the real roots discussed in Sec.~\ref{sec:roots}, we observe that for subkinks ($V<1$) this implies that the roots $\pm z_{2j}$, $j=1,\dots,l$, of $M_{+}(k)$ in ${\cal Z}_r^+$ and the roots $\pm p_{2j}$, $j=1,\dots,m$, of $M_{-}(k)$ in ${\cal P}_r^+$ contribute to $l^{+}(k)$, while the remaining roots $\pm z_{2j-1}$, $j=1,\dots,l+1$, of $M_{+}(k)$ in ${\cal Z}_r^-$ and $\pm p_{2j-1}$, $j=1,\dots,m+1$, of $M_{-}(k)$ in ${\cal P}_r^-$ contribute to $l^{-}(k)$. We thus obtain
\beq
\begin{split}
&l^{+}(k)=\dfrac{\prod\limits_{j=1}^{l}\left(1+\dfrac{(0 - ik)^2}{z_{2j}^2}\right)}{\prod\limits_{j=1}^{m}\left(1+\dfrac{(0 - ik)^2}{p_{2j}^2}\right)},\\
&l^{-}(k)
=\dfrac{\prod\limits_{j=1}^{l+1}\left(1+\dfrac{(0 + ik)^2}{z_{2j-1}^2}\right)}{\prod\limits_{j=1}^{m+1}\left(1+\dfrac{(0 + ik)^2}{p_{2j-1}^2}\right)}
\end{split}
\label{eq:lpm_subkinks}
\eeq
for subkinks. When $l=0$ or $m=0$, the corresponding products equal unity. Here we combined symmetric pairs $\pm r$ of real roots using
\[
\left(1-\dfrac{k}{r \pm i0}\right)\left(1-\dfrac{k}{-r \pm i0}\right)=1+\dfrac{(0\pm ik)^2}{r^2},
\]
where the notation $r \pm i0$ underscores the fact that the real roots are effectively shifted into the half-planes $\Im k \gtrless 0$. In particular, the zeroes and poles of $l^+(k)$ (the set ${\cal Z}_r^+ \cup {\cal P}_r^+$) are moved into $\Im k<0$, while the zeroes and poles of $l^-(k)$ (the set ${\cal Z}_r^- \cup {\cal P}_r^-$) are shifted into $\Im k>0$. In the case of shocks ($1<V<\gamma$) the sets ${\cal Z}_c^{\pm}$ are empty, and we have
\beq
\begin{split}
&l^{+}(k)=\dfrac{1}{\prod\limits_{j=1}^{m}\left(1+\dfrac{(0 - ik)^2}{p_{2j}^2}\right)},\\
&l^{-}(k)=\dfrac{1}{\prod\limits_{j=1}^{m+1}\left(1+\dfrac{(0 + ik)^2}{p_{2j-1}^2}\right)}.
\end{split}
\label{eq:lpm_shocks}
\eeq
Finally, in the superkink regime, both characteristic functions have no nonzero real roots, and thus we have
\beq
l^{\pm}(k)=1.
\label{eq:lpm_superkinks}
\eeq

We now consider the asymptotic behavior of the functions $L^\pm(k)$. Note first that equations \eqref{eq:Lfactor_more}-\eqref{eq:lpm_superkinks} imply that
\begin{equation}
L^\pm(k) \sim \sqrt{\frac{1-V^2}{\gamma^2-V^2}},\quad k\to \pm i0,
\label{eq:Asymptotics_Lpm_zeros}
\end{equation}
where we take the principal branch of the square root, which becomes purely imaginary when $1<V<\gamma$.
As shown in Appendix~\ref{sec:app}, the asymptotic behavior at infinity is given by
\begin{equation}
\begin{split}
&L^\pm(k) \sim R^{\mp 1},\quad k\to \pm i\infty, \quad \text{$V<1$ or $V>\gamma$},\\
&L^\pm (k) \sim R^{\mp 1}k^{\pm 1},\quad k \to \pm i\infty, \quad 1<V<\gamma
\end{split}
\label{eq:Asymptotics_Lpm_inf}
\end{equation}
where $R$ is given by
\begin{equation}
R=\dfrac{\Pi_z^+\Pi_p^-}{\Pi_z^-\Pi_p^+}=\dfrac{\prod\limits_{j=1}^{l} z_{2j}\prod\limits_{j=1}^{m+1} p_{2j-1}}{\prod\limits_{j=1}^{l+1} z_{2j-1}\prod\limits_{j=1}^{m} p_{2j}}, \quad V<1
\label{eq:R1}
\end{equation}
for subkinks,
\begin{equation}
R=\dfrac{\Pi_p^-}{\Pi_p^+}=\dfrac{\prod\limits_{j=1}^{m+1} p_{2j-1}}{\prod\limits_{j=1}^{m} p_{2j}}, \quad 1<V<\gamma
\label{eq:R2}
\end{equation}
for shocks, while for superkinks the absence of radiation implies
\[
R=1, \quad V>\gamma.
\]

Following the standard Wiener-Hopf procedure \cite{noble1958methods}, we perform the analytic continuation of both sides of Eq.~\eqref{eq:WH_second} to the entire complex plane and apply the Liouville theorem. Noting that the asymptotic estimates in Eq.~\eqref{eq:Asymptotics_Lpm_inf} imply that both sides of Eq.~\eqref{eq:WH_second} can be continued to a function that is at most linear in $k$, we obtain
\begin{equation}
\begin{split}
&L^+(k)\left[-\varepsilon_* -ik\left(\hat{\chi}^+(k)+\frac{\varepsilon_{+}}{0-ik}\right)\right]
\\
&=\frac{1}{L^-(k)}\left[-\varepsilon_*+ik\left(\hat{\chi}^-(k)+\frac{\varepsilon_{-}}{0+ik}\right)\right]\\
&=\psi_0+\psi_1 k.
\label{eq:WH_third}
\end{split}
\end{equation}
Here the constants $\psi_0$ and $\psi_1$ depend on the velocity regime due to the different asymptotic behavior in Eq.~\eqref{eq:Asymptotics_Lpm_inf} for kinks and shocks. Taking the limit $k \to \pm i0$ in Eq.~\eqref{eq:WH_third} and using the asymptotics  Eq.~\eqref{eq:Asymptotics_Lpm_zeros}, we obtain
\beq
\sqrt{\dfrac{1-V^2}{\gamma^2-V^2}}(\e_{+}-\e_*)=\sqrt{\dfrac{\gamma^2-V^2}{1-V^2}}(\e_{-}-\e_*)=\psi_0.
\label{eq:zero_limit}
\eeq
These relations hold for all velocities. Recalling Eq.~\eqref{eq:eps_star}, one can see that the first equality in Eq.~\eqref{eq:zero_limit} implies that the RH condition \eqref{eq:RH_PL} automatically holds for $\e_\pm$.

Observe now that by Eq.~\eqref{eq:Asymptotics_Lpm_inf}, both sides of the first equality in Eq.~\eqref{eq:WH_third} are constant at infinity when either $V<1$ or $V>\gamma$. Therefore, we must set $\psi_1=0$ in these velocity ranges. For subkinks ($V<1$) and superkinks ($V>\gamma$), taking the limits of the two sides of the first equality in Eq.~\eqref{eq:WH_third} as $k \to i\infty$ and $k \to -i\infty$, respectively, equating them to $\psi_0$ and applying the consistency condition \eqref{eq:consistency}, which implies $\lim_{s \rightarrow \infty} (s\hat{\e}^{\pm}(\pm is))=\e(0\pm)=\e_c$, then yields
\beq
\psi_0=\dfrac{\varepsilon_c-\varepsilon_*}{R},
\label{eq:psi0_kinks}
\eeq
where we recall Eq.~\eqref{eq:chi0}. Here $R$ is defined in Eq.~\eqref{eq:R1} for subkinks and $R=1$ for superkinks.
Equations \eqref{eq:zero_limit} and \eqref{eq:psi0_kinks} then imply that in these regimes
the limiting states $\varepsilon_{\pm}$ are fully determined by the velocity $V$ via
\begin{equation}
\varepsilon_{\pm}=\varepsilon_*+\frac{\varepsilon_c-\varepsilon_*}{R}\left(\frac{1-V^2}{\gamma^2-V^2}\right)^{\mp 1/2}.
\label{eq:eps_pm_kinks}
\end{equation}

Shocks ($1<V<\gamma$) correspond to the  generic case when both constants  $\psi_0$ and  $\psi_1$ in Eq.~\eqref{eq:WH_third} are nonzero. In this case the zero-limit equation \eqref{eq:zero_limit}, which still holds, and the limits $k \to \pm i\infty$ yield
\beq
 \psi_0=
i\sqrt{\frac{\gamma^2-V^2}{V^2-1}}\left(\varepsilon_--\varepsilon_*\right), \quad
\psi_1=\frac{\varepsilon_c-\varepsilon_*}{R},
\label{eq:psi01_shock}
\eeq
where $R$ is defined in Eq.~\eqref{eq:R2}.
Note, however, that although, as noted above, the RH condition \eqref{eq:RH_PL} is automatically satisfied for all three types of fronts,
in the case of shocks the limiting states $\varepsilon_{\pm}$ are \emph{not} uniquely determined by $V$, i.e., there is no condition that is equivalent to Eq.~\eqref{eq:eps_pm_kinks} we have for subkinks and superkinks. Therefore,  in the case of shocks one of the limiting states remains a free parameter, which agrees with the conclusions we reached while  considering the problem in both continuum and QC frameworks.

The solutions of the two equations in Eq.~\eqref{eq:WH_third} thus take the form
\begin{equation}
\hat{\chi}^\pm(k)=\frac{\varepsilon_*-\varepsilon_{\pm}}{0\mp ik}+\frac{\psi_0+\psi_1 k}{0\mp ik}\left[L^\pm(k)\right]^{\mp 1}.
\label{eq:Asymptotics_Solution_Fourier}
\end{equation}
Here $\psi_0$ is given by Eq.~\eqref{eq:psi0_kinks} and $\psi_1=0$ in the case of  both kinks and subkinks.  Instead, in the case of shocks  $\psi_0$ and $\psi_1$ are  given by Eq.~\eqref{eq:psi01_shock}. This yields the strains in the physical space given by
\begin{equation}
\varepsilon(\eta)=\varepsilon_{\pm} +\frac{1}{2\pi}\int_{-\infty}^{\infty}\hat{\chi}^\pm(k)e^{-ik\eta}\,dk,\quad \eta \gtrless 0,
\label{eq:Solution}
\end{equation}
where the integrals are computed by closing the contour of integration in $\Im k \lessgtr 0$ for $\eta \gtrless 0$ and applying the residue theorem. Here we recall that all real zeroes and singularities have been effectively shifted off the real axis into the corresponding half-planes.
As in the QC case, the solution can be then expressed in the general form \eqref{eq:soln_form}. Recall that this form includes localized ($\Phi_{\pm}(\eta)$) and radiative ($\Lambda_{\pm}(\eta)$) components.

The localized components $\Phi_{\pm}(\eta)$ are given by exponentially decaying functions arranged in the infinite sums
\beq
\begin{split}
&\Phi_{+}(\eta)=\sum_{z\in{\cal Z}_c^+}\frac{\omega_{-}^2(z)-(zV)^2}{2z^2V(\omega_+'(z)-V)}L^-(z)(\psi_0+\psi_1 z)e^{-iz\eta},\\
&\Phi_{-}(\eta)=\sum_{p\in{\cal P}_c^-}\frac{\omega_+^2(p)-(pV)^2}{2p^2V(\omega_{-}'(p)-V)}\frac{(\psi_0+\psi_1 p)}{L^+(p)}e^{-ip\eta}.
\end{split}
\label{eq:strain_Phi}
\eeq
The summation is over the sets of complex roots ${\cal P}_c^-$ (the poles of $L^-(k)$ in $\Im k>0$) and ${\cal Z}_c^+$ (the poles of $1/L^{+}(k)$ in $\Im k<0$) defined in \eqref{eq:root_sets}. To compute the residues we   used Eq.~\eqref{eq:L_def} and the identities $1/L^+(k)=L^{-}(k)/L(k)$ and $L^-(k)=L(k)/L^+(k)$ that follow from Eq.~\eqref{eq:Lfactor}.

The radiative components $\Lambda_{\pm}(\eta)$ in Eq.~\eqref{eq:soln_form} describe the lattice waves taking the energy from the moving front to infinity.
For subkinks ($V<1$), we have
\beq
 \begin{split}
 &\Lambda_{-}(\eta)=2\sum\limits_{j=1}^{m+1} \alpha_j^-\cos{(p_{2j-1}\eta+\beta_j^-)},\\
 &\Lambda_{+}(\eta)=2\sum\limits_{j=1}^{l} \alpha_j^+\cos{(z_{2j}\eta+\beta_j^+)},
 \end{split}
\label{eq:strain_rad}
\eeq
where the second sum is zero when $l=0$.  For shocks ($1<V<\gamma$), there is no radiation ahead of the front, so $\Lambda_{+}(\eta) \equiv 0$, while $\Lambda_{-}$ has the same form as above. The real coefficients $\alpha_j^\pm$ and $\beta_j^\pm$ can be obtained from the polar representation
 \[
 \begin{split}
 \alpha_j^+e^{-i\beta_j^+}&=\frac{L^-(z_{2j})\left[\omega_{-}^2(z_{2j})-(z_{2j}V)^2\right]}{2z_{2j}^2 V\left[\omega_+'(z_{2j})-V\right]}(\psi_0+\psi_1 z_{2j}),\\
\alpha_j^-e^{-i\beta_j^-}&=-\frac{\left[\omega_+^2(p_{2j-1})-(p_{2j-1}V)^2\right]}{2p_{2j-1}^2 V\left[V-\omega_{-}'(p_{2j-1})\right]L^+(p_{2j-1})}\\
&\times (\psi_0+\psi_1 p_{2j-1})
\end{split}
\]
with the corresponding values of $\psi_0$ and $\psi_1$. Only the second equation is relevant for shocks since $\Lambda_+=0$ in that case.  Here we used Eq.~\eqref{eq:sing_plus} and Eq.~\eqref{eq:sing_minus} obtained in Appendix~\ref{sec:app}. Finally, for superkinks ($V>\gamma$) there is no radiation either ahead or behind the propagating front, and so in this case $\Lambda_{-}(\eta)=\Lambda_{+}(\eta) \equiv 0$.

In addition to strains we can also explicitly compute the particle velocities $v(\eta)$. To this end we need to solve the equation   $v(\eta+1)-v(\eta)=-V\varepsilon'(\eta)$, where $\e(\eta)$ is given by Eq.~\eqref{eq:soln_form}, Eq.~\eqref{eq:strain_Phi} and Eq.~\eqref{eq:strain_rad}. Using Fourier transform, we obtain
\[
v(\eta)=v_{\pm}+\Theta_{\pm}(\eta)+\Upsilon_{\pm}(\eta), \quad \eta\gtrless 1/2,
\]
where $v_+-v_{-}=-V(\e_+-\e_-)$ coincides with the first RH condition in Eq.~\eqref{eq:RHconds} for the continuum problem, and since one of $v_\pm$ is arbitrary by Galilean invariance, we may set $v_{\pm}=-V\e_{\pm}$. Here we can also identify the exponentially decaying terms
\[
\begin{split}
\Upsilon_{+}&=-\sum_{z\in{\cal Z}_c^+}\frac{\omega_{-}(z)-(zV)^2}{4z\sin{\frac{z}{2}}[\omega_+'(z)-V]}L^-(z)\\
 & \times (\psi_0+\psi_1 z)e^{iz(\eta-1/2)},\\
\Upsilon_{-}&=-\sum_{p\in{\cal P}_c^-}\frac{\omega_+(p)-(pV)^2}{4p\sin{\frac{p}{2}}[\omega_{-}'(p)-V]}\frac{(\psi_0+\psi_1 p)}{L^+(p)}e^{ip(\eta-1/2)}
\end{split}
\]
and the oscillatory terms $\Theta_{\pm}(\eta)$ describing  radiation. For subkinks ($V<1$), we have
\beq
\begin{split}
&\Theta_{+}(\eta)=-\sum_{j=1}^{l}\frac{\alpha_j^{+}z_{2j}V}{\sin\frac{z_{2j}}{2}}\cos{(z_{2j}(\eta-1/2)+\beta_j^+)},\\
&\Theta_{-}(\eta)=-\sum_{j=1}^{m+1}\frac{\alpha_j^{-}p_{2j-1}V}{\sin\frac{p_{2j-1}}{2}}\cos{(p_{2j-1}(\eta-1/2)+\beta_j^-)},\\
\end{split}
\label{eq:vel_rad}
\eeq
where the second sum is zero when $l=0$.  For shocks ($1<V<\gamma$), the function $\Theta_{-}(\eta)$ has the same form, while  $\Theta_{+}(\eta) \equiv 0$. For superkinks, $\Theta_{-}(\eta)=\Theta_{+}(\eta) \equiv 0$.

\subsection{Dissipation rate}
The knowledge of the exact solution of the discrete problem  gives us the access to the energy (phonon) radiation from the moving fronts to infinity. As we have already mentioned, since the radiated  energy is lost by the  front,    the associated rate of the energy transport to infinity by lattice waves   can be interpreted as the rate of  dissipation.

Following the procedure we used for the QC model, we again consider the cumulative energy fluxes $G_+$ and $G_-$ emitted ahead and behind the front. Recalling Eq.~\eqref{eq:Rpm_gen}, we find that dissipation rates $R_{\pm}=G_{\pm}V$ on both sides are zero for superkinks, which involve no phonon radiation, and thus $G_{+}=G_{-}=0$ in this case. For subkinks ($V<1$) we obtain
\[
\begin{gathered}
{\cal R}_+ = \sum_{j=1}^l\langle {\cal E}_+(z_{2j})\rangle(\omega_+'(z_{2j})-V),\\
{\cal R}_- = \sum_{j=1}^{m+1}\langle {\cal E}_{-}(p_{2j-1})\rangle(V-\omega_-'(p_{2j-1})),
\end{gathered}
\]
where ${\cal R}_+=0$ when $l=0$, and ${\cal E}_{+}(z_{2j}) = v_j^2/2 +\e_j^2/2$ and ${\cal E}_{-}(p_{2j-1}) = v_j^2/2 +\gamma^2\e_j^2$/2 are energy densities carried by individual lattice waves with (real and positive) wave numbers $z_{2j}\in {\cal Z}_r^+$ and $p_{2j-1}\in {\cal P}_r^-$, respectively, and the averaging is over the corresponding time periods. Using the expressions for strains $\e_j$ in Eq.~\eqref{eq:strain_rad} and particle velocities $v_j$ in Eq.~\eqref{eq:vel_rad} of the emitted waves with the corresponding wave numbers, we obtain
\begin{equation}
\begin{gathered}
G_+=2\sum_{j=1}^l(\alpha_j^+)^2\omega_+^2(z_{2j})\left(\dfrac{\omega_+'(z_{2j})}{V}-1\right),\\
G_-=2\gamma^2\sum_{j=1}^{m+1}(\alpha_j^-)^2\omega_{-}^2(p_{2j-1})\left(1-\dfrac{\omega_{-}'(p_{2j-1})}{V}\right),
\end{gathered}
\end{equation}
where $G_+=0$ when $l=0$. For shocks ($1<V<\gamma$), $G_{-}$ has the same form, and $G_{+}=0$. This yields explicit expressions for the driving force $G=G_{+}+G_{-}$ in different velocity regimes. Alternatively, we can compute the driving force
from the macroscopic area-difference formula \eqref{eq:Dissipation} (with $E_1=1$ and $E_2=\gamma^2$ in the dimensionless formulation). Using Eq.~\eqref{eq:eps_pm_kinks} for the kink regimes, Eq.~\eqref{eq:RH_PL} for shocks and recalling Eq.~\eqref{eq:eps_star}, we obtain
\[
G=\begin{cases}\frac{\gamma^2-1}{2}\left(1-\frac{1}{R^2}\right)\left(\varepsilon_c-\varepsilon_*\right)^2, & V<1,\\
\frac{\gamma^2-1}{2}[\left(\varepsilon_c-\varepsilon_*\right)^2+\frac{V^2-1}{\gamma^2-V^2}\left(\varepsilon_+-\varepsilon_*\right)^2],& V \in (1,\gamma),\\
0, & V>\gamma.
\end{cases}
\]
For subkinks and superkinks this yields the kinetic relations $G=G(V)$ (recall that $R$ depends on $V$ via Eq.~\eqref{eq:R1} in the subkink regime), which complement the classical RH conditions, while for shocks the driving force  remains  dependent on the choice of $\e_+$, which, as we recall, is a free parameter in this case. We have verified that these `macroscopic'  expressions for $G$ are equivalent to the ones obtained by computing directly the energy fluxes.

\subsection{Admissibility}
\begin{figure}[h!]
  \centering
 \center{\includegraphics[scale=0.4]{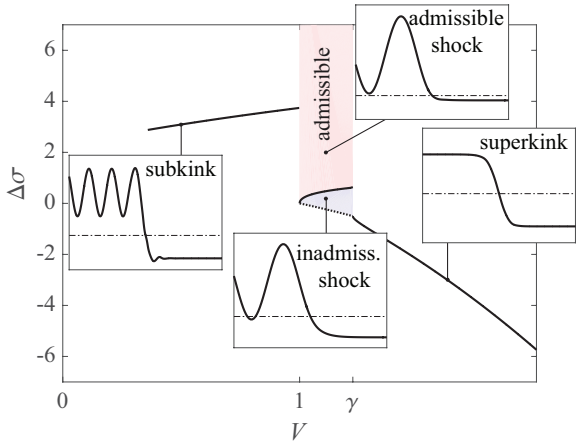}}
  \caption{ Admissibility sets of solutions of the discrete problem. In the blue region we observe $\varepsilon(\eta)\leq \varepsilon_c$ when $\eta<\varepsilon_c$, and the dashed lower boundary of the region marks the threshold $\varepsilon_{-}=\varepsilon_c$. The insets show examples of the strains $\varepsilon(\eta)$. Here $\gamma^2=1.5$, $\varepsilon_c=1$, and
we set $\varepsilon_{+}=0$.}
\label{fig:admissibilityD}
\end{figure}
As in the case of the QC approximation, one still needs to verify which of  the obtained solutions are admissible, i.e., satisfy Eq.~\eqref{eq:admissibility}.  In Fig.~\ref{fig:admissibilityD} we show  the admissibility diagram for the discrete problem, which is a direct analog of the similar diagram for the QC model presented in Fig.~\ref{fig:admissibilityQC}. As in that case, admissible subkink and shock solutions in the discrete problem feature a single radiation mode propagating behind the front, where the wave number $r$ is a positive root of the characteristic equation $\omega_{-}(r)=Vr$, while $\Lambda_{+}\equiv 0$. In the superkink case, $\Lambda_{\pm}\equiv 0$. In the case of shocks  one of the limiting states remains a free parameter, which agrees with both continuum and QC approximations. One can see that for $V<1$ sufficiently fast subkinks are admissible. For $V>\gamma$, all superkinks satisfy the assumed inequalities. In the interval $1<V<\gamma$ the TW solutions describing shock waves are admissible inside the pink domain. In the blue domain such TW solutions are not admissible and are replaced by the DSWs, as we will discuss in the next subsection.

We conclude that the main features of the QC regime diagram Fig.~\ref{fig:admissibilityQC} are preserved in the full discrete model. Thus, both types of  kinks, represented in  Fig.~\ref{fig:admissibilityD} by one dimensional manifolds, are admissible (for sufficiently large $V$ in the case of subkinks). Shocks are again not defined uniquely for a given $\Delta\sigma$ and are admissible for sufficiently large values of $\Delta\sigma$. The two diagrams differ significantly only at small $V<1$, where the QC model, as expected, does not capture the complex resonant behavior of the (typically inadmissible) slow discrete subkinks.

Our comparison suggests that outside the regimes of particularly slow subkinks, all three types of transition fronts are adequately described  by only few roots of the characteristic equation capturing  long (but not infinitely long) lattice waves. This implies that carefully designed QC theories with only few parameters (describing the crucial mesoscopic scales) can be successful in capturing such  a fundamental nonlinear dynamic effect as radiative friction. It also points to the paramount importance of the QC reproduction of the relevant mesoscopic time scales, in addition to the more conventional task of modeling the internal length scales. In other words, the task of the adequate dispersive approximation of the kinetic energy may be at least as challenging    as  the task of the satisfactory representation of the nonlocal elastic energy.

\subsection{Numerical simulations}
To test the stability of the obtained analytical solutions, we conducted a series of numerical simulations, in which, starting with Riemann initial data, we  traced the emergence of the nonlinear transition fronts propagating at constant velocity. More specifically, we solved numerically the system \eqref{eq:EoM_strain} (rescaled so that $\rho=1$ and $h=1$) with $N=1000$ springs and discontinuous initial conditions of the form
\[
\varepsilon_n(0)=\begin{cases}
\varepsilon_l,\,n<500,\\
0,\,n\geq 500,
\end{cases}\quad \frac{d\varepsilon_n}{dt}(0)=0
\]
and free boundary conditions. We used the Dormand-Prince algorithm (ode45 in Matlab), and the duration of simulations was such that the boundaries did not affect the front dynamics.
In each simulation we varied $\varepsilon_l$ and $\Delta\sigma$, while keeping all other parameters fixed. As in the case of QC model, we identified  four generic types of traveling fronts  which all emerged and stabilized by the numerical time $t=500$.
\begin{figure}[h!]
  \centering
  \begin{tabular}{@{}p{0.45 \linewidth}@{\quad}p{0.45 \linewidth}@{}}
    \subfigimg[width=\linewidth]{(a)}{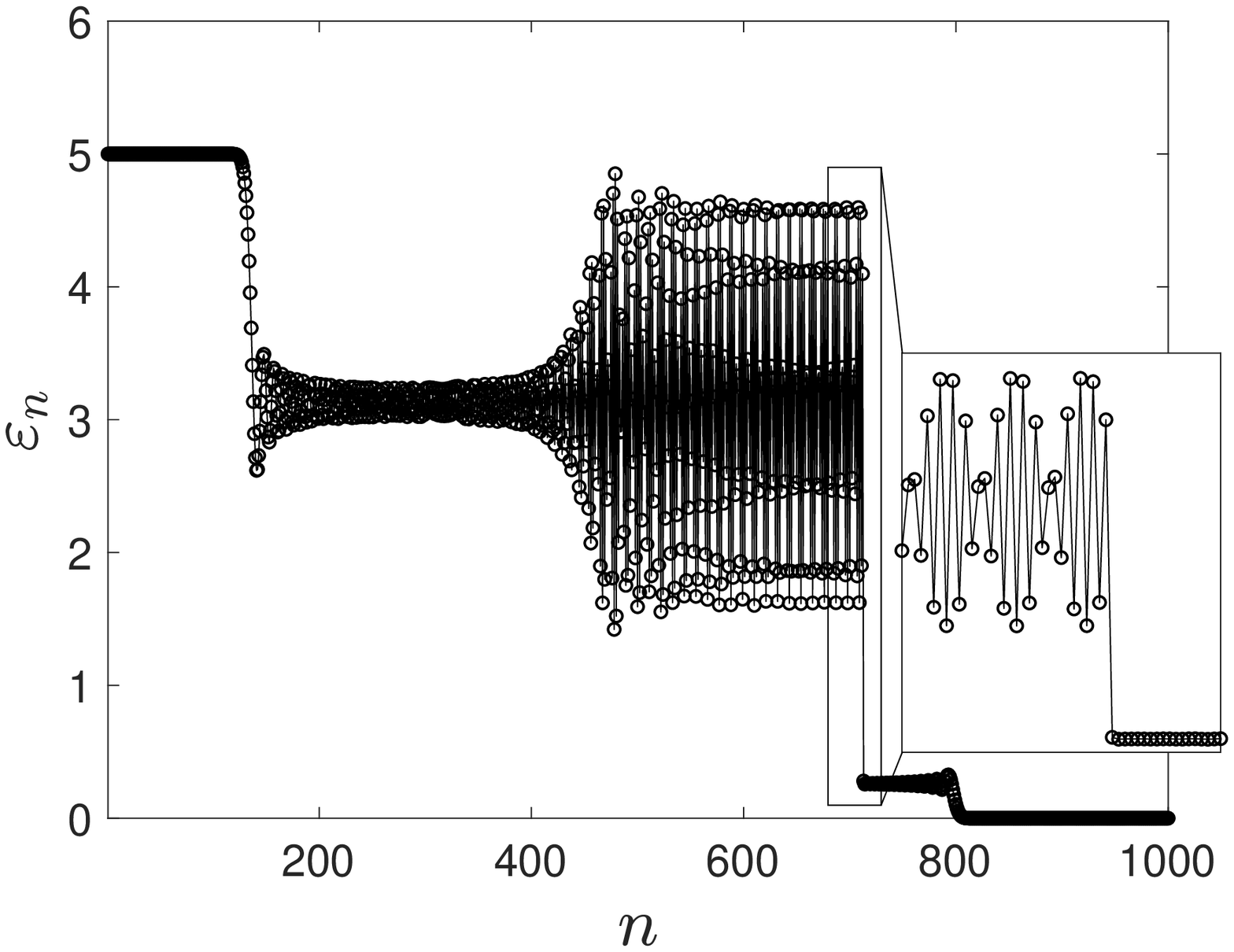} &
    \subfigimg[width=\linewidth]{(b)}{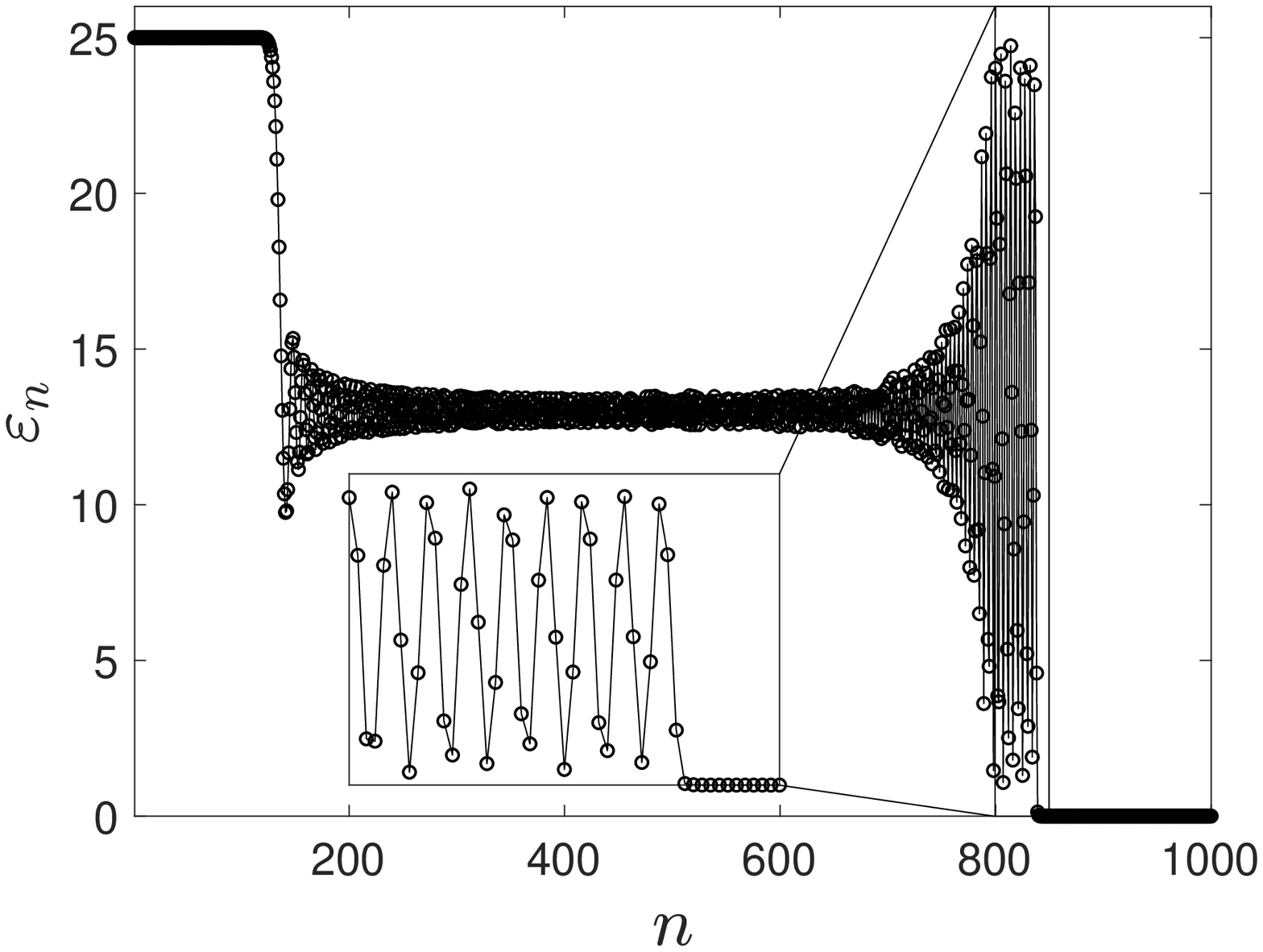} \\
    \subfigimg[width=\linewidth]{(c)}{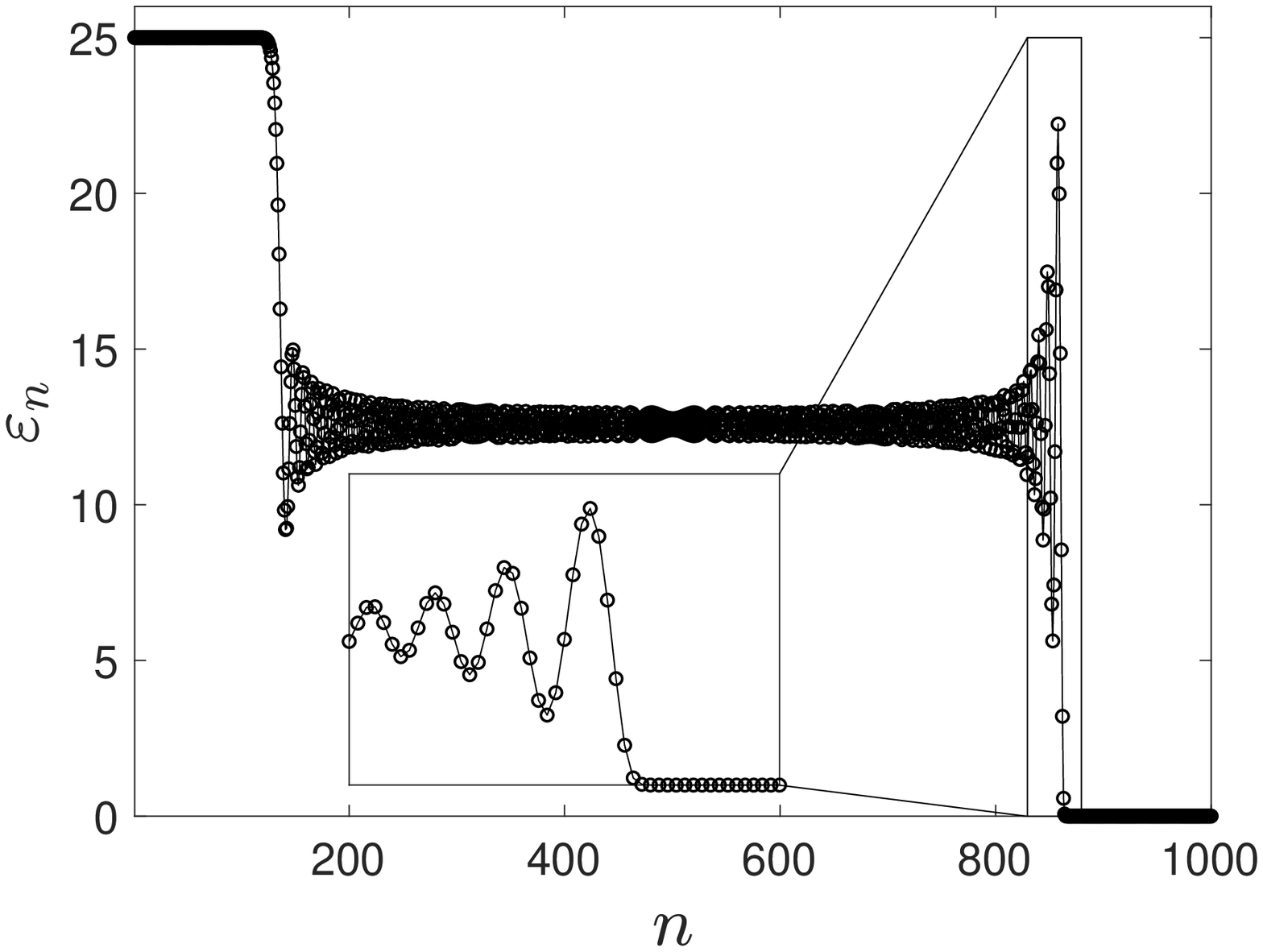} &
    \subfigimg[width=\linewidth]{(d)}{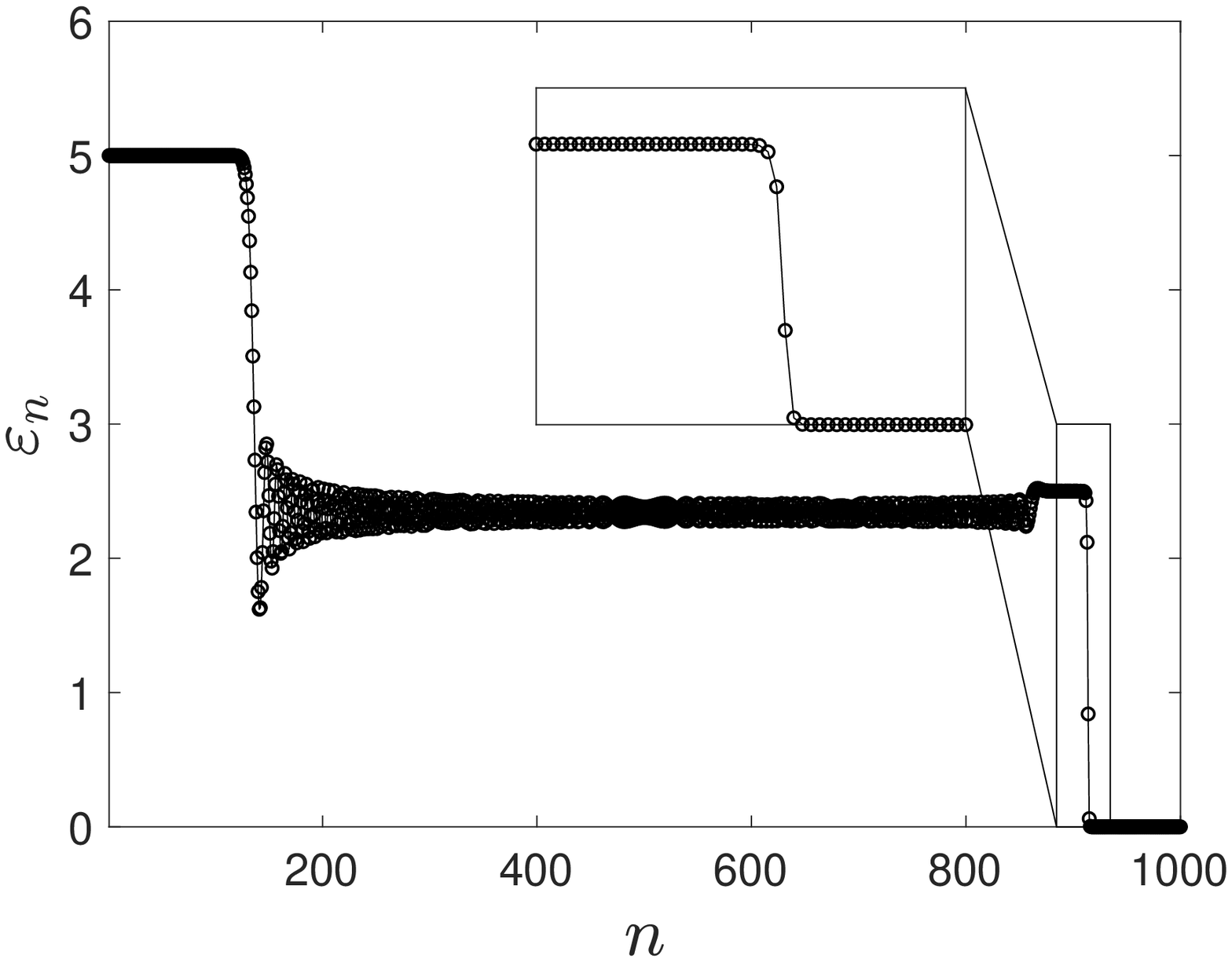}
  \end{tabular}
  \caption{ Different regimes of front propagation in FPU chain under Riemann-type initial conditions with different left strain $\varepsilon_l$ and $\Delta\sigma$: (a) subkink ($\varepsilon_l=5$, $\Delta\sigma=2.5$); (b) conventional shock ($\varepsilon_l=25$, $\Delta\sigma=2.5$); (c) dispersive shock ($\varepsilon_l=25$,  $\Delta\sigma=0$); (d) superkink ($\varepsilon_l=5$, $\Delta\sigma=-1.5$). Here $\gamma^2=1.5$, $\varepsilon_c=1$ and $t=300$.}
\label{fig:numerics}
\end{figure}

The results of the simulations are summarized in Fig.~\ref{fig:numerics}. They confirm the possibility of stable propagation of all three types of transition waves. For all steady transition fronts, the  localized  waveforms are accompanied by linear dispersive waves appearing behind the transition front and moving away from it with velocity $-\gamma$.  In the case of a subkink shown in Fig.~2(a), there is also a linear dispersive wave propagating ahead of the transition front with velocity $1$.

Our results suggest stability of the all three regimes, subkinks, shocks and superkinks inside the corresponding  admissible domains of the  $(V,\Delta\sigma)$ plane. Recall that subkinks are admissible when $V<1$ is sufficiently large. An example of a subkink propagation is shown in Fig.~\ref{fig:numerics}(a). We found that superkinks can only appear when $V>1$ and $\Delta\sigma<-\varepsilon_c(\gamma^2-1)<0$. An example is shown in Fig.~\ref{fig:numerics}(d). Recall also that shocks are only admissible when $1<V<\gamma$ and $\Delta\sigma$ is above a certain threshold, as shown in Fig.~\ref{fig:admissibilityD}. An example of an admissible shock propagation is shown in Fig.~\ref{fig:numerics}(b). Inside the domain of inadmissible shocks  we expectedly do not find  steady transition fronts but find instead the spreading transition profiles of DSW type (Fig.~\ref{fig:numerics}(c)), similar to the corresponding prediction of the QC model. We reiterate that the DSWs are mentioned here only for completeness. The detailed study of such non-steady regimes is outside the scope of this paper, not in the least because these solutions are well documented in the literature. They appear here naturally as  stable replacements for the inadmissible traveling waves.

\section{Applications in  metamaterial design}
\label{Applications to metamaterials}
The importance  of metamaterials is due to their ability to exploit post-instability structural responses.  Effectively,   metamaterials  utilize internal changes in the sub-elements, which imitate molecular phase transitions at supermolecular scales.  The  success of metamaterial paradigm  is due to the fact that artificial `meta-molecules' with desired properties can be manufactured at the relevant scales.

The  localized transition fronts, studied in this paper, can be viewed as elementary bites of mechanical information that can be generated, delivered and erased in periodic lattice  metamaterials. Due to the presence of stress-sensitive repeating structural units, such metamaterials can   manipulate mechanical information  using  advantageously the dispersion of elastic waves. By  carefully tailoring  relationships between characteristic dimensions, one can design metamaterials combining  the effects of   strong dispersion with various  forms of energy nonconvexity.  One of the main challenges in the design of  metamaterial structures is to ensure that the switching takes place at a predefined levels of stress and that the particular switching waves are  generated  when the task is, for instance, to enhance actuation or  perform energy harvesting.

In view of these and other  potential applications, the prototypical FPU model, studied in this paper,  can serve  as a proof of concept showing the broad  variety of the functionally  distinct   switching regimes which can be controlled by the deliberate parameter tuning. Even though the actual  3D  metamaterials with the desired  properties  would still have to be designed,  the  results obtained in this paper already now provide a specific guidance regarding, for instance,  which metamaterial  should be chosen to ensure a supersonic, dissipation-free, communication of mechanical information, as opposed to a design favoring subsonic switching which ensures a heavily dissipative response.

\section{Conclusions}
\label{Conclusions}
The goal of this paper was to reveal the interrelations between structurally different steadily moving transition (switching) fronts in the classical FPU model. Our main result is the demonstration that this non-integrable Hamiltonian model supports three   classes of such   fronts that can be classified  as subsonic (subkinks), intersonic (shocks) and supersonic (superkinks).

To obtain analytical results we limited our analysis to piecewise linear elastic responses.  In this case exact solutions of the discrete model for each class of fronts can be presented in the form of infinite series. Within this setting, we have shown that the proposed classification is exhaustive. The common framework considered in this work allows us to describe all  three types of switching waves in a unified way and associate distinct types of switching waves with particular classes of  elastic responses.

While  the constructed explicit solutions  of the discrete problem  are sufficient to corroborate  these qualitative claims, the origin of the difference between the three types of fronts remains relatively opaque in the FPU setting  dealing with an infinite system of nonlinear ordinary differential equations. To achieve conceptual  transparency,  we constructed a QC approximation of the FPU problem. An excellent  agreement with the behavior of the discrete model was obtained using a long-wave (infra-red) approximation utilizing only two internal scales. We stress that the successful  coarse-grained theory relies  on the approximation of kinetic  energy, in contrast to more conventional asymptotic approaches  such as the KdV model and its higher order analogs.

A detailed comparison of the exact solutions for the  QC theory and the discrete problem showed that the chosen approximation adequately describes the complex interrelation between all three types of the transition fronts.  This means that the whole complexity of the dispersive structure of the FPU model was not really necessary for the description of the main features of these special solutions. In other words, the dispersive properties of all three different classes of fronts can be satisfactorily captured using a simple QC model.

Our analysis also reveals that the obtained macroscopically dissipative front profiles, describing subkinks and shocks, cannot be adequately described by the continuum nonlinear wave equation, as may be suggested by a naive homogenization. Instead, they should be interpreted as microscopic descriptions of Whitham shocks connecting oscillatory and constant states \cite{sprenger2020discontinuous,gavrilyuk2020stationary}. Such generalized (dispersive) shocks usually correspond to heteroclinic traveling waves of the a dispersive model connecting standard critical points with periodic orbits. To capture such connections in a PDE format we had to use a higher order QC model.

To fully understand the different structure of the three types of transition fronts, we have drawn upon a broad variety of physical and mathematical considerations, including characteristics, barriers, topological transitions, undercompressive nature, critical manifolds and kinetic relations, which all point to the existence of exactly three universality classes of transition fronts. In this sense the obtained perspective  can be viewed as unifying not only for the description of switching waves but also for different analytical approaches to the analysis of nonlinear dispersive systems.

Several important issues have been naturally left for future studies. The traveling wave description of the switching waves is clearly incomplete when it comes to transient effects like interaction with obstacles and multiple collisions. The approach to such problems proposed for special cases in \cite{truskinovsky2010beyond} can be also generalized and applied in our more general framework. The present work does not address thermal effects, which may become relevant for metamaterial with submicron scale mimicking cytoskeleton or extracellular environment. For these purposes the approach proposed in  \cite{ngan2002thermo} can be generalized here as well. Another issue that we have not addressed in this work concerns different modes of manipulation and control of transition fronts from a distance using DC and AC-type dynamic loading, which is of particular interest for metamaterial applications. The successful use of such control has been recently demonstrated for semilinear discrete systems in \cite{gorbushin2020frictionless}.\\

\noindent {\bf Acknowledgments.} The authors thank G. Mishuris for helpful discussions. The work of AV was supported by the NSF grant DMS-1808956. LT and NG acknowledge the support of the French Agence Nationale de la Recherche under the grant ANR-17-CE08-0047-02.

\appendix

\section{Some asymptotic results}
\label{sec:app}
To obtain the asymptotic behavior at infinity, we follow \cite{slepyan1982antiplane} and observe that for subkinks ($V<1$) we have
\beq
\begin{split}
&H_z^{\pm}(k)=\sqrt{1-V^2}\prod_{k \in {\cal Z}_c^{\pm}}\biggl(1-\dfrac{k}{z}\biggr)=\\
&\dfrac{V\Pi_z^{+}\Pi_z^{-}}{T_z^{\pm}(k)}
\exp\biggr[\pm\dfrac{1}{2\pi i}\int\limits_{-\infty}^{\infty}\ln \biggl(\dfrac{H_z(\xi)T_z(\xi)}{V^2\Pi_z^2}\biggr)\dfrac{d\xi}{\xi-k\mp i0}\biggr]\\
&\sim V\Pi_z^{+}\Pi_z^{-}(0\mp ik)^{-2l-1}, \quad k \to \pm i\infty,
\end{split}
\label{eq:Hz_sub}
\eeq
where
\[
T_z^{\pm}(k)=(0 \mp ik)^{2l+1}, \quad T_z=T_z^{+}T_z^{-}, \quad H_z=H_z^{+}H_z^{-}
\]
and
\[
\Pi_z^+=\prod\limits_{j=1}^{l} z_{2j}, \quad \Pi_z^-=\prod\limits_{j=1}^{l+1} z_{2j-1}, \quad \Pi_z=\Pi_z^{+}\Pi_z^{-},
\]
and
\beq
\begin{split}
&H_p^{\pm}(k)=\sqrt{\gamma^2-V^2}\prod_{k \in {\cal P}_c^{\pm}}\biggl(1-\dfrac{p}{z}\biggr)=\\
&\dfrac{V\Pi_p^{+}\Pi_p^{-}}{T_p^{\pm}(k)}
\exp\biggr[\pm\dfrac{1}{2\pi i}\int\limits_{-\infty}^{\infty}\ln \biggl(\dfrac{H_p(\xi)T_p(\xi)}{V^2\Pi_p^2}\biggr)\dfrac{d\xi}{\xi-k\mp i0}\biggr]\\
&\sim V\Pi_p^{+}\Pi_p^{-}(0\mp ik)^{-2m-1}, \quad k \to \pm i\infty,
\end{split}
\label{eq:Hp}
\eeq
where
\[
T_p^{\pm}(k)=(0 \mp ik)^{2m+1}, \quad T_p=T_p^{+}T_p^{-}, \quad H_p=H_p^{+}H_p^{-}
\]
and
\[
\Pi_p^+=\prod\limits_{j=1}^{m} p_{2j}, \quad \Pi_p^-=\prod\limits_{j=1}^{m+1} p_{2j-1}, \quad \Pi_p=\Pi_p^{+}\Pi_p^{-}.
\]
Here the expressions under the logarithms in the Cauchy-type factorization integrals are set up in such a way that they tend to $1$ as $k \to \pm i\infty$, while the logarithms remain real along the entire integration path \cite{slepyan1982antiplane}. These asymptotic expressions imply that in the subkink regime
\[
L_0^{\pm}(k)=\dfrac{H_z^{\pm}(k)}{H_p^{\pm}(k)} \sim \dfrac{\Pi_z^{+}\Pi_z^{-}}{\Pi_p^{+}\Pi_p^{-}}(0\mp ik)^{2(m-l)}, \quad k \to \pm i\infty,
\]
while
\[
l^{\pm}(k) \sim \dfrac{(\Pi_p^{\pm})^2}{(\Pi_z^{\pm})^2}(0 \mp ik)^{2(l-m)}, \quad k \to \pm i\infty,
\]
so that
\[
L^{\pm}(k) \sim R^{\mp 1}, \quad k \to \pm i\infty, \quad V<1,
\]
where $R$ is given by Eq.~\eqref{eq:R1}.
For shocks ($1<V<\gamma$) Eq.~\eqref{eq:Hp} still holds but due to the absence of nonzero real roots of $M_{+}(k)$ in this regime, Eq.~\eqref{eq:Hz_sub} is replaced by \cite{slepyan1982antiplane}
\beq
\begin{split}
&H_z^{\pm}(k)=i\sqrt{V^2-1}\prod_{k \in {\cal Z}_c^{\pm}}\biggl(1-\dfrac{k}{z}\biggr)=\\
&iV\exp\biggr[\pm\dfrac{1}{2\pi i}\int\limits_{-\infty}^{\infty}\ln \biggl(\dfrac{H_z(\xi)}{V^2}\biggr)\dfrac{d\xi}{\xi-k\mp i0}\biggr]\\
&\sim iV, \quad k \to \pm i\infty,
\end{split}
\label{eq:Hz_super}
\eeq
so that
\[
L_0^{\pm}(k) \sim \dfrac{i}{\Pi_p^{+}\Pi_p^{-}}(0\mp ik)^{2m+1}, \quad k \to \pm i\infty,
\]
which together with
\[
\begin{split}
&l^{+}(k) \sim (\Pi_p^{+})^2(0 - ik)^{-2m}, \quad k \to i\infty\\
&l^{-}(k) \sim (\Pi_p^{-})^2(0 + ik)^{-2(m+1)}, \quad k \to -i\infty,
\end{split}
\]
implies that
\[
L^{\pm} \sim R^{\mp 1}k^{\pm 1}, \quad k \to \pm i\infty, \quad 1<V<\gamma,
\]
where $R$ is given by Eq.~\eqref{eq:R2}.
Finally, for superkinks ($V>\gamma$), both characteristic functions have no nonzero real roots, and thus $H_z^{\pm} \sim iV$ as in Eq.~\eqref{eq:Hz_super} and $H_p^{\pm} \sim iV$ in the limit $k \to \pm i\infty$. Together with \eqref{eq:lpm_superkinks} this implies $L^{\pm} \sim 1$ as $k \to \pm i\infty$ in this velocity regime. Combining these results, we obtain Eq.~\eqref{eq:Asymptotics_Lpm_inf}.

Recalling Eq.~\eqref{eq:L_def} and Eq.~\eqref{eq:Lfactor}, one can also show that near the real singularities
\beq
\begin{split}
&\frac{1}{L^+(k)} \sim \frac{\omega_{-}^2(z_{2j})-(z_{2j}V)^2}{{2z_{2j}Vi |\omega'_+(z_{2j})-V|}}\\
& \times \frac{L^-(z_{2j})}{0-i(k-z_{2j})}, \quad k\to z_{2j},
\end{split}
\label{eq:sing_plus}
\eeq
and
\beq
\begin{split}
&L^-(k) \sim \frac{\omega_{+}^2(p_{2j-1})-(p_{2j-1}V)^2}{{2p_{2j-1}Vi |\omega'_-(p_{2j-1})-V|}}\\
& \times \frac{1}{L^+(p_{2j-1})}\frac{1}{0+i(k-p_{2j-1})}, \quad k\to p_{2j-1},
\end{split}
\label{eq:sing_minus}
\eeq
with similar expressions for the negative real singular points.


%

\end{document}